\documentclass[11pt,a4paper]{article}
\usepackage{graphicx} 
\usepackage{amsmath}
\usepackage{amssymb}
\usepackage{booktabs}
\usepackage{cite}
\usepackage[colorlinks=true]{hyperref}
\usepackage{colortbl}
\usepackage[margin=1in]{geometry}
\usepackage{authblk}
\usepackage{lipsum}
\usepackage{adjustbox}
\usepackage{placeins}
\usepackage{cleveref}
\usepackage{chngcntr}
\usepackage{xcolor}
\usepackage{soul}
\newcommand{\bpm}[0]{\begin{pmatrix}}
\newcommand{\epm}[0]{\end{pmatrix}}

\def\Acknowledgements{\bigskip  \bigskip \begin{center} \begin{large}
             \bf Acknowledgements\end{large}\end{center}}

\title{The impact of CP-violating phases on DM observables in the cpMSSM}
\author[1]{Melissa van Beekveld \thanks{mbeekvel@nikhef.nl}}
\author[2]{Wim Beenakker \thanks{w.beenakker@science.ru.nl}}
\author[2]{Jochem Kip \thanks{jochem.kip@ru.nl}}
\author[2,3]{Marrit Schutten}
\author[2]{Dirren van Vlijmen \thanks{dirren.vanvlijmen@ru.nl}}

\affil[1]{Nikhef, Theory Group, Science Park 105, 1098 XG, Amsterdam, The Netherlands}
\affil[2]{Institute for Mathematics, Astrophysics and Particle Physics, Radboud University Nijmegen, Heyendaalseweg 135, Nijmegen, the Netherlands}
\affil[3]{Van Swinderen Institute for Particle Physics and Gravity, University of Groningen, 9747 AG Groningen, The Netherlands}

\begin{document}

\maketitle
\abstract{In this paper we examine the effect of adding CP-violating phases to the pMSSM on the Dark-Matter (DM) direct-detection cross sections, the velocity-weighted annihilation cross section, and the DM relic density. We show that $\varphi_{M_2}$ and $\varphi_{\mu}$, the phases of the wino and higgsino mass parameters, are sufficiently constrained by electron electric-dipole-moment (EDM) measurements such that the allowed values of these phases are too small to influence these DM observables. Conversely, the phase of the bino-mass parameter $\varphi_{M_1}$ and the phases associated with the trilinear couplings, $\varphi_{A^t}$, $\varphi_{A^b}$, and $\varphi_{A^\tau}$, are barely constrained by EDM experiments. We find that these cpMSSM phases can have an important impact on the mentioned DM observables. This especially concerns pMSSM points that lie on the boundary of exclusion, as observables can be affected by cpMSSM phases to the extent that they are either pushed into the observable region, or outside the excluded region.}
\newpage
\section{Introduction}
The electric dipole moment (EDM) of the electron provides one of the most stringent limits on new sources of CP violation. The ACME collaboration has reported an upper bound of $1.1\cdot10^{-29}\,\, {\rm e\, cm}$ on the electron EDM~\cite{ACME:Andreev2018}, and more recently the JILA experiment has improved this bound to $4.1\cdot 10^{-30}\,\,{\rm e\, cm}$~\cite{JILA:_EDM:Roussy:2022cmp}. A new limit is expected by the eEDM-NL collaboration~\cite{Groningen_eEDM:Aggarwal_2018}. However, these constraints are still far removed from the Standard Model (SM) prediction for the electron EDM, which lies at approximately $10^{-38}\,\,{\rm e\,cm}$~\cite{SM_eEDM:Pospelov:2005pr, SM_eEDM2:Yamaguchi:2020eub, SM_eEDM3:Yamaguchi:2020dsy}. Notably, this value does not include contributions of the PMNS matrix, which is expected to increase the SM prediction of the electron EDM when included~\cite{eEDM_PMNS:Ng:1995cs}. Regardless, the gap between the theoretical and experimental values provides ample space for new sources of CP violation. In fact, new beyond the SM sources of CP violation are required in order to explain the baryon asymmetry in the Universe, which is currently measured to be $4.7\cdot 10^{-10} < \eta_{B} < 6.5\cdot 10^{-10}$~\cite{BAU1:Planck:2015fie, BAU2:Planck:2018vyg, BAU3:WMAP:2008lyn, BAU4:WMAP:2003elm}. The SM alone lacks sufficient CP violation to explain this asymmetry and must therefore be supplemented with new CP-violating physics~\cite{CP_SM_lacking:Cline:2006ts}. This makes CP violation one of the more promising areas to find signs of physics beyond the SM.\\
Another strong sign of new physics is of course the absence of a suitable Dark Matter (DM) candidate in the SM. One model that simultaneously provides a DM candidate and allows for additional sources of CP violation is the MSSM~\cite{SM_eEDM:Pospelov:2005pr, eEDM_diagrams:John_Ellis_2008, MSSM_CP3:Pilaftsis:2002fe, MSSM_CP2, MSSM_CP4:Li:2010ax}. In the MSSM the couplings between SM particles and their superpartners depend in part on the composition of said sparticles~\cite{Primer:Martin:1997ns, Sparticles:drees2004theory}. Thus by changing the mixing matrices of the sparticles, the coupling strengths of the MSSM will change. Especially the relative real and imaginary components of sparticle mixing matrices are influenced by the addition of phases. Therefore changes in observables that are not explicitly dependent on CP violation are to be expected when adding CP-violating phases.\\
In this paper we will discuss the impact of CP-violating phases on the spin-dependent and spin-independent cross sections, $\sigma_{p/n}^{SD}$ and $\sigma_{p/n}^{SI}$ respectively, the velocity-weighted DM annihilation cross section $\langle \sigma v \rangle$, and the DM relic density $\Omega h^2$. A number of previous studies have investigated such dependencies~\cite{Micr_CP:Belanger:2006qa, MSSM_CP_DM:Abe:2018qlw, MSSM_CP_DM1:Belanger:2006pc, MSSM_CP_DM2:Lee:2007ai, MSSM_CP_DM3:Belanger:2009dz, MSSM_CP_DM4:Belanger:2008yc, MSSM_CP_DM5:Kraml:2007pr}, however these studies are either performed in a SUGRA-like model, make simplifying assumptions, or have been performed before the discovery of the Higgs boson. Here we will use the phenomenological MSSM~\cite{pMSSM:MSSMWorkingGroup:1998fiq} (pMSSM) with its 19 free real parameters as a baseline and allow for complex parameters, resulting in additional phases. We call this model with a total of 25 free parameters the cpMSSM. \\
This paper is structured as follows. We start with discussing the effect of adding CP-violating phases to the Higgs, electroweakino, and sfermion sectors in the MSSM in Section~\ref{sec:Theory}. In Section~\ref{sec:Methodology} we specify the computational software we have used and discuss our sampling method. We then continue in Section~\ref{sec:Results} with a discussion on the impact of the phases on various DM observables. We finish with our conclusions in Section~\ref{sec:Conclusion}.
\section{Theory}
\label{sec:Theory}
When allowing for complex parameters in the pMSSM, 9 phases are added in total, of which 6 are independent. There are 8 pMSSM parameters that obtain a phase in the cpMSSM: $M_1$, $M_2$, $M_3$, $\mu$, $A^t$, $A^b$, $A^\tau$, $B_\mu$. These are the gaugino mass parameters, Higgs-mixing mass term in the superpotential, trilinear couplings between sfermions and Higgs bosons, and the SUSY-breaking Higgs-mixing mass term respectively. Furthermore, a new phase $\eta$ between the two Higgs doublets is introduced~\cite{Higgs_phase:HEINEMEYER2007300}. When including the phase between the Higgs doublets $H_u$ and $H_d$, the expansion of the neutral fields around the vevs $v_u$ and $v_d$ reads
\begin{align}
     H_u = e^{i\eta}\bpm H^+_u \\ H^0_u \epm  = \frac{e^{i\eta}}{\sqrt{2}}\bpm \sqrt{2} H_u^+ \\ v_u + \phi_u+i\sigma_u \epm\,, && H_d = \bpm H^0_d \\ H^-_d \epm = \frac{1}{\sqrt{2}} \bpm v_d + \phi_d + i\sigma_d \\ \sqrt{2}H_d^-\epm\,.
\end{align}
Here the phase $\eta$ has been assigned to $H_u$ by convention; by using an appropriate $U(1)_Y$ gauge transformation $\eta$ can be freely shifted between $H_u$ and $H_d$.\\
The Higgs potential is minimized for the vevs of the fields as $\langle H_u^0 \rangle =v_u/\sqrt{2}$, $\langle H_d^0\rangle = v_d/\sqrt{2}$, $\langle H^+_u\rangle = 0$, and $\langle H^-_d \rangle = 0$, and is encoded in the tadpole equations of the neutral fields. In the cpMSSM an additional non-trivial tadpole equation is present as a result of mixing real and imaginary component fields of the Higgs doublets. The complete set of tadpole equations in the cpMSSM is given by~\cite{CP_Tadpoles:PILAFTSIS199888}
\begin{subequations}
\begin{align}
    \frac{1}{v_u}\frac{\partial V}{\partial \phi_u} = |\mu|^2 + m_{H_u}^2 - |B_\mu|\cot(\beta)\cos(\varphi_b+\eta) - \frac{1}{2}m_Z^2\cos(2\beta) + \frac{1}{v_u}\frac{\partial V^{\text{corr}}}{\partial \phi_u} = 0\,,\\
    \frac{1}{v_d}\frac{\partial V}{\partial \phi_d} = |\mu|^2 + m_{H_d}^2 - |B_\mu|\tan(\beta)\cos(\varphi_b+\eta) + \frac{1}{2}m_Z^2\cos(2\beta) + \frac{1}{v_d} \frac{\partial V^{\text{corr}}}{\partial \phi_d} = 0\,,\\
    \frac{1}{v_u}\frac{\partial V}{\partial \sigma_u} = \sin(\varphi_b +\eta)|B_\mu| + \frac{1}{v_d}\frac{\partial V^{\text{corr}}}{\partial \sigma_u} = \frac{1}{v_d}\frac{\partial V}{\partial \sigma_d} = \sin(\varphi_b +\eta)|B_\mu| + \frac{1}{v_u}\frac{\partial V^{\text{corr}}}{\partial \sigma_d} = 0\,.    \label{eq:tadpole eqations}
\end{align}
\end{subequations}
Here $V$ is the Higgs potential, $V^{\text{corr}}$ are higher-order corrections to the tree-level Higgs potential, $\varphi_b$ is the phase of $B_\mu$, $m_{H_u}^2$ and $m_{H_d}^2$ are the SUSY-breaking Higgs mass parameters, $\tan(\beta)$ is the ratio between the Higgs vevs $v_u/v_d$, and $m_Z$ is the mass of the $Z$ boson. In our analysis we use these tadpole equations to derive the values of $|\mu|^2$, $B_\mu$ and $\eta$ using $m_{H_u}^2$, $m_{H_d}^2$, $\tan\beta$, and those parameters required for the calculation of $V^{\text{corr}}$ as input. \\
The corrections to the Higgs potential are given by one-particle-irreducible vacuum diagrams~\cite{Tadpole_vacuum_diagrams_1:Martin:2001vx, Tadpole_vacuum_diagrams_2:PhysRevD.9.1686}. The top-stop sector typically provides the largest corrections, since the largest corrections typically arise from $h_1$-mediated diagrams, which couples SM-like. However, contributions from the bottom-sbottom and tau-stau sectors can become sizable for large values of $\tan(\beta)$ if the masses of $h_2$ and $h_3$ are not exceedingly large. In the following we give all diagrams with mediating stops; the sbottom and stau diagrams can be obtained by simply interchanging $\tilde{t}_{1,2}$ with $\tilde{b}_{1,2}$ or $\tilde{\tau}_{1,2}$. The top-stop-mediated one-loop vacuum diagrams are:
\begin{align}
    \adjustbox{valign = c}{\includegraphics{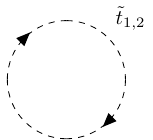}}
    &&\adjustbox{valign = c}{\includegraphics{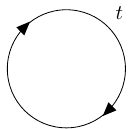}}\,,\label{eq:one loop tadpole correction diagrams}
\end{align}
which clearly only depend on the masses of the top quark $t$ and stop mass eigenstates $\tilde{t}_{1,2}$. The relevant two-loop Feynman diagrams instead have a dependence on not only the particle masses, but also on the mixing of the particles and their couplings. Some important diagrams are:
\begin{align}
    &  \adjustbox{valign = c}{\includegraphics{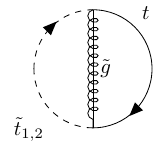}}
    && \adjustbox{valign = c}{\includegraphics{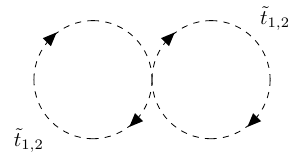}}
    &&  \adjustbox{valign = c}{\includegraphics{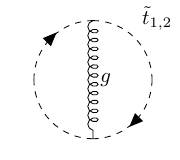}} \nonumber \\
    & \adjustbox{valign = c}{\includegraphics{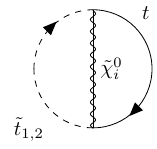}}
    &&  \adjustbox{valign = c}{\includegraphics{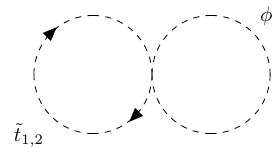}}
    &&  \adjustbox{valign = c}{\includegraphics{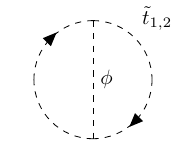}} \label{eq: two loop tadpole corrections diagrams}
\end{align}
Here $\phi$ indicates any Higgs boson, $\tilde{\chi}^0_i$ the neutralinos, $\tilde{g}$ the gluino, and $g$ the gluon. Explicit expressions for the corrections to the tadpole equations can be found in Refs.\cite{Tadpole_corrections1:Baer_2013, Tadpole_corrections2:DEDES2003333}. The diagrams of Figures~\eqref{eq:one loop tadpole correction diagrams}~and~\eqref{eq: two loop tadpole corrections diagrams} are not exhaustive, but are the most relevant ones for our purposes, as will be seen in Section~\ref{subsec:The electron EDM limit and implications for the phases}. For example, the stop-sbottom diagram with a mediating chargino is not shown. We emphasize that the last diagram of Figure~\eqref{eq: two loop tadpole corrections diagrams} has an explicit dependence on the trilinear coupling between the sfermions and Higgs boson. \\
When inspecting the tadpole equations at tree level, i.e.~$V^{\text{corr}}=0$, we can see that $\varphi_b = -\eta$, i.e.~at tree level the Higgs sector is CP-conserving. However, higher-order corrections can induce a discrepancy between $\varphi_b$ and $\eta$, thereby introducing CP violation in the Higgs sector. Consequently, the neutral mass eigenstates as known from the pMSSM, the CP-even $h^0$ and $H^0$ and CP-odd $A^0$ Higgs bosons, undergo mixing into three Higgs bosons $h_1$, $h_2$ and $h_3$~\cite{Higgs_cp_mixing:Pilaftsis_1999}. We parameterize the mixing of the three neutral Higgs CP-conserved mass eigenstates into the three new mass-ordered mass eigenstates with a matrix $\mathcal{R}$
\begin{align}
    \begin{pmatrix} h_1 \\ h_2 \\ h_3 \end{pmatrix} = \begin{pmatrix} \mathcal{R}_{1h^0} & \mathcal{R}_{1H^0} & \mathcal{R}_{1A^0} \\ \mathcal{R}_{2h^0} & \mathcal{R}_{2H^0} & \mathcal{R}_{2A^0} \\ \mathcal{R}_{3h^0} & \mathcal{R}_{3H^0} & \mathcal{R}_{3A^0}\end{pmatrix} \begin{pmatrix} h^0 \\ H^0 \\ A^0 \end{pmatrix}\,. \label{eq: Higgs mixing}
\end{align}
In the neutralino sector the mass parameters $M_1 = |M_1|e^{i\varphi_{M_1}}$, $M_2 = |M_2|e^{i\varphi_{M_2}}$, and $\mu = |\mu|e^{i\varphi_{\mu}}$ all obtain a phase, which in addition to the phase $\eta$ between the Higgs doublets, provides four phases in the neutralino mass matrix, and three in the chargino mass matrix. The neutralino mass matrix in the basis $\bpm \tilde{B}, & \tilde{W}, & \tilde{h}_d^0, & \tilde{h}_u^0\epm$ reads
\begin{align}
    m_{\tilde{\chi}^0} = \bpm
        M_1                         & 0                     & -g^\prime v_d/2   & g^\prime v_u e^{-i\eta}/2 \\
        0                           & M_2                   & g_2 v_d/2         & -g_2 v_u e^{-i\eta}/2      \\
        -g^\prime v_d/2             & g_2 v_d/2             & 0                 & -\mu                      \\
        g^\prime v_u e^{-i\eta}/2   & -g_2 v_ue^{-i\eta}/2  & -\mu              & 0                         
    \epm \,.
\end{align}
Here $g^\prime$ is the $U(1)_Y$ associated coupling constant, and $g_2$ that of $SU(2)_L$. Furthermore, $\tilde{B}$, $\tilde{W}$, $\tilde{h}_d^0$, and $\tilde{h}_u^0$ are the bino, wino, and neutral higgsino down and up-type fields respectively. The neutralino mass matrix is diagonalized by a unitary matrix $N$ as $N^*m_{\tilde{\chi}^0}N^\dagger = M^D_{\tilde{\chi}^0}$ with $M^D_{\tilde{\chi}^0}$ being a diagonal matrix with real entries. The mixing matrix $N$ provides the amount of bino $|N_{i1}|$, wino $|N_{i2}|$, or higgsino $\sqrt{|N_{i3}|^2 + |N_{i4}|^2}$ component of neutralino $\tilde{\chi}^0_i$.\\
In the chargino sector the mass matrix $M_{\tilde{C}}$, in the basis $\bpm \tilde{W}^+, & \tilde{h}_u^+, & \tilde{W}^-, & \tilde{h}_d^-\epm$ is given by
\begin{align}
    M_{\tilde{C}} = \bpm 
    0                   & 0                             & M_2                           & g_2v_d/\sqrt{2}   \\
    0                   & 0                             & g_2 v_u e^{-i\eta}/\sqrt{2}   & \mu               \\
    M_2                 & g_2 v_u e^{-i\eta}/\sqrt{2}   & 0                             & 0                 \\
    g_2 v_d /\sqrt{2}   & \mu                           & 0                             & 0                 
    \epm
    \equiv \bpm 0 & m_{\tilde{\chi}^\pm}^T \\ m_{\tilde{\chi}^\pm} & 0 \epm\,.
\end{align}
We diagonalize $m_{\tilde{\chi}^\pm}$ with two unitary matrices $\mathcal{U}$ and $\mathcal{V}$ such that $\mathcal{U}^*m_{\tilde{\chi}^\pm}\mathcal{V}^\dagger$ is diagonal with real entries. In the chargino sector both $M_2$ and $\mu$ obtain a phase, as well as all $v_u$ entries, which obtain a factor $e^{-i\eta}$. \\
Lastly, the sfermion sector obtains a phase for the trilinear couplings $A^f$, $\mu$, and again factors $e^{\pm i\eta}$ from the Higgs sector. For illustration purposes we explicitly show the mass matrix for the staus, which in the $\bpm \tilde{\tau}_L, & \tilde{\tau}_R\epm$ basis reads
\begin{align}
    M^2_{\tilde{\tau}} = \bpm
    M^2_{\tilde{\tau}_L}+M_Z^2(T^{\tilde{\tau}}_{3L}-Q_\tau\sin^2(\theta_W))\cos(2\beta)+m_\tau^2  & m_\tau (A^{\tau^*}-\mu e^{i\eta}\tan(\beta)) \\
    m_\tau(A^\tau - \mu^* e^{-i\eta}\tan(\beta)  )                                        & M^2_{\tilde{\tau}_R} - \sin^2(\theta_W)M_Z^2\cos(2\beta) + m_\tau^2
    \epm\,, \label{eq:sfermion mass matrix}
\end{align}
with $M_{\tilde{\tau}_L}^2$ and $M_{\tilde{\tau}_R}^2$ the left and right-handed stau mass terms, $A^\tau$ the relevant trilinear coupling, $Q_\tau$ the charge of the tau, $T^{\tilde{\tau}}_{3L}$ the third component of the weak isospin of $\tilde{\tau}_L$, and $m_\tau$ the tau mass. The mass matrices for the stop and sbottom sectors are defined analogously, and can be found in Ref.~\cite{Sparticles:drees2004theory}. Notably, the mass matrices for the first two generations are similar, but do not include the trilinear couplings, as we assume them to be zero in the pMSSM. Furthermore, in the pMSSM we assume that there is no generational mixing, which allows us to diagonalize the sfermion mass matrices in terms of $2\times2$ blocks. For the third generation of sfermions the diagonalization matrices are $X^{\tilde{t}}$, $X^{\tilde{b}}$, and $X^{\tilde{\tau}}$ for the stop, sbottom, and stau sectors respectively.\\
The MSSM Lagrangian has two accidental $U(1)$ symmetries that can be used to rotate two phases away. We shall name these $U(1)_A$ and $U(1)_B$, which are equivalent to the $R$ and Peccei-Quin symmetries~\cite{U1AB-1:PhysRevD.16.1791, U1AB-2:PhysRevLett.38.1440, U1AB-3:GABBIANI1996321}. If we assign the MSSM component fields the charges $Q_A$ and $Q_B$ of Table~\ref{tab:U1AB charges MSSM} such that
\begin{align}
    \psi \overset{U(1)_A}{\longrightarrow} \psi' = e^{iQ_A \omega_A} \psi\,, && \psi \overset{U(1)_B}{\longrightarrow} \psi' e^{iQ_B\omega_B} \psi\,,
\end{align}
and transform the following parameters as
\begin{align}
    &\mu \rightarrow e^{2i(\omega_A + \omega_B)} \mu\,, && B_\mu \rightarrow e^{4i\omega_A}B_\mu\,, \nonumber \\
    &M_{1,2,3} \rightarrow e^{2i(\omega_A - \omega_B)} M_{1,2,3}\,, && A^{t,b,\tau} \rightarrow  e^{2i(\omega_A - \omega_B)}A^{t,b,\tau}\,, \label{eq:U(1) rotations}
\end{align}
then the MSSM Lagrangian is left invariant. We use the third tadpole equation of Eq.~\eqref{eq:tadpole eqations} in conjunction with the $U(1)_A$ and $U(1)_B$ symmetries to set $\eta=0$ such that the vevs are real, thereby avoiding the additional factors $e^{i\eta}$ in the neutralino, chargino, and sfermion mass matrices. Furthermore, we set $\varphi_{M_3}=0$, seeing as this phase appears in two-loop order corrections at the earliest for the observables we consider.\\
\begin{table}[t]
    \centering
    {\renewcommand{\arraystretch}{1.3}
    \begin{tabular}{cccccc}
    \rowcolor[gray]{.6}
    Sparticles & $Q_A$ & $Q_B$ & Particles & $Q_A$ & $Q_B$\\
    sfermions   & 0     & \phantom{-}1  & fermions      & \phantom{-}1  & 0 \\
    \rowcolor[gray]{.9}
    higgsinos   & -1    & -1            & Higgs bosons  & -2            & 0\\
    gauginos    & -1    & \phantom{-}1  & gauge bosons  & \phantom{-}0  & 0
    \end{tabular}
    }
    \caption{The $U(1)_A$ and $U(1)_B$ charges of the MSSM component fields. These charges are given for the left-handed fields, thus the right-handed fermion and sfermion fields instead have charges $-Q_A$ and $-Q_B$.}
    \label{tab:U1AB charges MSSM}
\end{table}
We compute the electron EDM with all one-loop corrections and the dominant two-loop contributions, which are given by the Barr-Zee diagrams. At one loop there are two contributing diagrams, mediated by a neutralino--selectron and chargino--electron-sneutrino pair
\begin{align}
    \adjustbox{valign=b}{\includegraphics{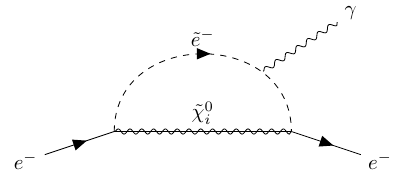}}\,,\quad
    \adjustbox{valign=b}{\includegraphics{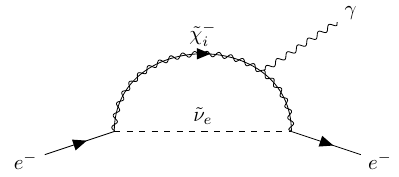}}\,. \label{eq:1_loop_eEDM_diagrams}
\end{align}
At two-loop order we only take the dominant diagrams into account, which are given by the following Barr-Zee diagrams:
\begin{align}\allowdisplaybreaks
    \adjustbox{valign=b}{\includegraphics{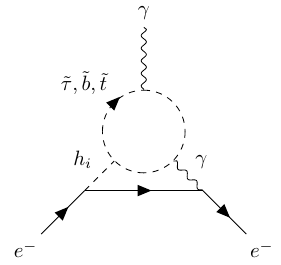}}\,, \quad 
    \adjustbox{valign=b}{\includegraphics{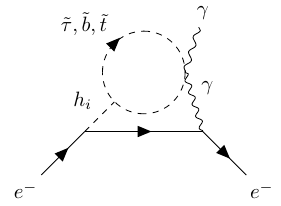}}\,, \nonumber \\
    \adjustbox{valign=b}{\includegraphics{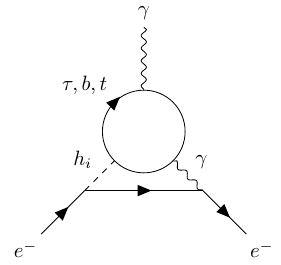}}\,, \quad 
    \adjustbox{valign=b}{\includegraphics{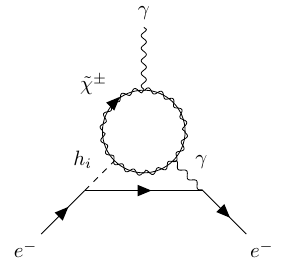}}\,. \label{eq:2_loop_eEDM_diagrams}
\end{align}
The expressions for these diagrams can be found in Ref.~\cite{eEDM_diagrams:John_Ellis_2008, Thesis_Marrit}. Notably, additional Barr-Zee diagrams with charged Higgs contributions exist, but these have been shown to be subdominant, providing at most a $\mathcal{O}(10\%)$ contribution\footnote{These expressions are evaluated in the Feynman-'t Hooft gauge.}~\cite{Charged_barr-zee:Pilaftsis:1999td}. Moreover, there are additional diagrams, collectively known as rainbow diagrams, that can potentially become as large as the Barr-Zee contributions for $\mathcal{O}(TeV)$-scale sparticle masses~\cite{rainbow_diagrams:Yamanaka:2012ia}. We do not take these diagrams into account as we expect these contributions to be subleading for our considered model points.
\section{Methodology}
\label{sec:Methodology}
\subsection{Spectrum generation}
We compute the spectrum of a cpMSSM model point (a set of input parameters) using a custom-made version of {\tt SPheno 4.0.4}~\cite{SPheno_1:POROD2003275, SPheno_2:POROD20122458} which is built with model files\footnote{The files can be found in Ref.~\cite{Thesis_Marrit}} generated by {\tt SARAH}~\cite{SARAH_1:STAUB20141773, SARAH_2:STAUB20122165, SARAH_3:Goodsell_2015, SARAH_4:Goodsell_2017}. We subsequently use {\tt FeynHiggs 2.18.1}~\cite{FeynHiggs_1:Heinemeyer_2007, FeynHiggs_2:Hollik_2014, FeynHiggs_3Hollik_2014, FeynHiggs_4:Bahl_2020, FeynHiggs_5:bahl2019precision, FeynHiggs_6:Bahl_2018, FeynHiggs_7:Bahl_2016, FeynHiggs_8:Hahn_2014, FeynHiggs_9:Frank_2007, FeynHiggs_10:Degrassi_2003, FeynHiggs_11:Heinemeyer_1999, FeynHiggs_12:Heinemeyer_2000} for an increased precision in the computation of the Higgs masses. The DM relic density, direct detection cross sections, and velocity-weighted cross section are computed with {\tt Micromegas 5.2.13}~\cite{Micromegas_1:B_langer_2006, Micromegas_2:BELANGER2007367, Micromegas_3:BELANGER2009747, Micromegas_4:BELANGER2011842, Micromegas_5:BELANGER2014960, Micromegas_6:B_langer_2021, Micromegas_7:B_langer_2006}. The electron EDM is determined with an in-house code using the Feynman diagrams of Eqs.~\eqref{eq:1_loop_eEDM_diagrams} and \eqref{eq:2_loop_eEDM_diagrams} and their concomitant expressions~\cite{eEDM_diagrams:John_Ellis_2008}. Our imposed limits are specified in Section~\ref{subsec:Limits}.\\
We note that {\tt FeynHiggs} implicitly assumes that $B_\mu$ is real, while we have chosen to set $\eta=0$. While this poses no issue at tree level, loop effects may induce a mismatch between $\varphi_b$ and $\eta$. We remedy this by using the identities of the $U(1)_A$ and $U(1)_B$ symmetries given in Eq.~\eqref{eq:U(1) rotations} in order to perform a rotation such that $B_\mu$ is real for the {\tt FeynHiggs} computation.  After the computation performed by {\tt FeynHiggs} we rotate all phases back such that $\eta=0$ in accordance with our chosen convention. An extensive overview of the various existing conventions regarding notation and computation, including ours, can be found in Ref.~\cite{Thesis_Marrit}.\\
To sample model points we start by using pMSSM model points to which we add phases. We use the pMSSM parameters as input by using both the sign and absolute value of their values. A cpMSSM parameter will thus for example be made as $\text{sign}(M_1)|M_1|e^{i\varphi_{M_1}}$, where $M_1$ is the pMSSM parameter and $\varphi_{M_1}$ its phase that must be added in the cpMSSM. We construct our initial set of pMSSM model points by both adapting the dataset from~\cite{gm2:VanBeekveld:2021tgn} and by uniformly scanning the pMSSM parameter space for valid pMSSM model points such that we have pMSSM model points over a wide range of sparticle masses and compositions. Specifically, we consider masses of the lightest neutralino up to 2 TeV, and have searched for valid low-mass solutions for the stau, stop, and sbottom. Especially the sfermion masses of the third generation become relevant in the Barr-Zee diagrams of the electron EDM, c.f.~Ref.~\cite{eEDM_diagrams:John_Ellis_2008}.\\
We use pMSSM model points from Ref.~\cite{gm2:VanBeekveld:2021tgn} since that study optimised for values of the muon anomalous magnetic moment $(g-2)_\mu$ which could explain the experimental and theoretical discrepancy reported at the time~\cite{gm2_FL2:Abi_2021}. The MSSM contributions to $(g-2)_\mu$ arise from different contributions of the same diagrams that provide electron EDM contributions~\cite{eEDMtheory:Nowakowski_2005}. In this study we use this discrepancy in addition to a $2\sigma$ uncertainty window only as an upper bound for $(g-2)_\mu$. As such, we expect pMSSM model points with a comparatively high $(g-2)_\mu$ value to represent model points for which the phases are likely to be the most constrained. Furthermore, we have scanned the pMSSM parameter space with a flat prior, such that we do not optimise for any particular observable. We do so to see the effect of adding phases on typical pMSSM model points. From the uniform scan of the pMSSM parameter space we have taken care to include model points that contain all relevant different processes for the various DM observables that we consider.\\
We subsequently add phases to the pMSSM model points to obtain cpMSSM model points. As such, a single pMSSM model point can give rise to a variety of different cpMSSM model points. We call such a pMSSM model point a pMSSM seed. We sample our phases between $-\pi/2$ and $\pi/2$. Notably, all parameters to which we add phases can have a minus sign in the pMSSM, which effectively corresponds to a phase of $\pi$. Therefore, we limit our sampling to $\pm \pi/2$ without affecting the physics. We sample the phases using a normalizing flow network~\cite{Normalizing_flows:rezende2016variational}, as opposed to a logarithmic prior in order to more efficiently sample the parameter space of the phases. Especially since we are interested in finding all relevant phenomenology, as opposed to exhaustively scanning the cpMSSM parameter space. The specifics are detailed in Subsection~\ref{subsec:scanning procedure}, and more extensively in Ref.~\cite{Thesis_Dirren}. The input parameters of both the pMSSM and cpMSSM are shown in Table~\ref{tab:input parameters}.\\
\begin{table}[t]
    \centering
    {\renewcommand{\arraystretch}{1.3}
    \begin{tabular}{cc}
    \rowcolor[gray]{.6}
    pMSSM & cpMSSM \\
    $M_1,\,M_2,\,M_3$,\,$\text{sign}(\mu)$ & $\varphi_{M_1},\,\varphi_{M_2},\,\varphi_{\mu}$ \\ 
    $A^t,\,A^b,\,A^\tau$ & $\varphi_{A^t},\,\varphi_{A^b},\,\varphi_{A^\tau}$ \\
    \rowcolor[gray]{.9}
    $\tan(\beta),\,m^2_{H_u},\,m^2_{H_d}$ & \\
    $m_{\tilde{Q}_1},\,m_{\tilde{u}_R},\,m_{\tilde{d}_R},\,m_{\tilde{L}_1},\,m_{\tilde{e}_R}$ & \\
    $m_{\tilde{Q}_3},\,m_{\tilde{t}_R},\,m_{\tilde{b}_R},\,m_{\tilde{L}_3},\,m_{\tilde{\tau}_R}$ \\ 
    \end{tabular}
    }
    \caption{The chosen input parameters of the pMSSM are on the left and the parameters that are added when transitioning to the cpMSSM are shown on the right. All pMSSM parameters are real. Note that $\text{sign}(\mu)$ is subsumed in $\varphi_{\mu}$ in the cpMSSM. A complete cpMSSM parameter is thus made as $\text{sign}(M_1)|M_1|e^{i\varphi_{M_1}}$.}
    \label{tab:input parameters}
\end{table}
We have opted to use {\tt SPheno} as opposed to {\tt CPsuperH}~\cite{CPsuperH_1:Lee_2004, CPsuperH_2:Lee_2009, CPsuperH_3:Lee_2013}, for more precise sparticle mass calculations. This is due to the fact that the computation of the sparticle masses is performed via RGE running in {\tt SPheno} and in an effective potential approach in {\tt CPsuperH}~\cite{CPsuperH_diff:Carena_2000}. These two approaches can differ significantly for identical input parameters when sparticle masses lie far apart. This is due to the fact that in the effective potential approach sparticle masses are computed at a single scale, thus corrections that arise from running the RGE's can become significant when sparticles differ greatly in mass. These differences are especially important in the stop sector, which provide significant corrections to the Higgs boson mass. Some examples of input parameters where such differences arise are provided in the appendix of Ref.~\cite{Thesis_Marrit}. 
\subsection{Limits}
\label{subsec:Limits}
For all of our model points we require that the lightest Higgs boson is between 122 and 128 GeV, allowing for a 3 GeV uncertainty on the computation of the Higgs mass, and its CP-odd component $\mathcal{R}_{1A^0}^2$ must not exceed 10\%~\cite{CP_Higgs_maximal_mixing}. Additionally, the mass of the second Higgs boson must be heavier than 350 GeV, such that the Higgs sector is in the decoupling limit~\cite{Higgs_decoupling:Bauer:2017ota}. The LEP limits dictating that the lightest chargino must be heavier than 103.5 GeV, the first two generation sleptons must be heavier than 90 GeV, and the staus must be above 85 GeV~\cite{LEP_limits1, LEP_limits2:200273} are also implemented. We run no detector simulation for our cpMSSM model points, due to a lack of suitable software. However, for our pMSSM seeds we run {\tt Prospino 2}~\cite{Prospino:Beenakker:1996ed} to compute the electroweakino production cross sections to take electroweakino limits into account. We subsequently run {\tt Smodels}~\cite{Smodels1:Ambrogi:2018ujg, Smodels2:Heisig:2018kfq, Smodels3:Dutta:2018ioj, Smodels4:Ambrogi:2017neo, Smodels5:Kraml:2013mwa} on the pMSSM seeds to quickly determine whether the pMSSM seed is excluded by the LHC searches. We note in passing that it would be interesting to perform a dedicated analysis regarding the impact of CP-violating phases in the cpMSSM on LHC signatures.
To be conservative, we additionally implement ATLAS limits from both compressed and uncompressed scenarios for sleptons and staus ~\cite{LHC_slepton_limits:ATLAS:2017vat, sleptons2:ATLAS:2019lff,sleptons3:ATLAS:2019lng, sleptons4:ATLAS:2022hbt}. The masses of the coloured sparticles are demanded to be above 2.5 TeV for the gluino, 1.2 TeV for the stop and sbottom, and 2 TeV for the remaining squarks, such that they evade limits from ATLAS and CMS.\\
Furthermore, we implement the spin-dependent, $\sigma_{p/n}^{SD}$, and spin-independent, $\sigma_{p/n}^{SI}$, limits of Xenon-1T~\cite{Xenon-1T:PhysRevLett.121.111302}, Xenon-nT~\cite{Xenon-nT:PhysRevLett.131.041003}, Pico-60~\cite{Pico60-1:PhysRevD.100.022001, Pico60-2:PhysRevLett.118.251301}, PandaX~\cite{PandaX-II:Wang_2020, PandaX-4T:PhysRevLett.127.261802}, and LZ~\cite{LZ:PhysRevLett.131.041002}. The DM relic density $\Omega h^2$ of our DM candidate, the lightest neutralino, must lie between 0.09 and 0.15~\cite{Relic_density:2020}, where we have included a 0.03 margin for uncertainties arising from the computation of the relic density. In addition, we implement the Fermi-LAT gamma-ray limits from dwarf-spheroidal galaxies~\cite{FERMI-LAT:Ackermann_2015}, and the H.E.S.S. limits from the Galactic centre~\cite{HESS:Abdalla_2022}. Lastly, the electron EDM must be below $4.1\cdot 10^{-30}$~e~cm~\cite{JILA:_EDM:Roussy:2022cmp}.
\subsection{Scanning procedure}
\label{subsec:scanning procedure}
To sample CP-violating phases for a given pMSSM model point we have trained a conditional normalizing flow model~\cite{Normalizing_flows:rezende2016variational}. Such a model transforms an initial distribution into a target distribution. It does so by using a set of invertible transformations with a known Jacobian. Such transformations are known as bijectors. In our case the initial distribution is a 7-dimensional Gaussian distribution\footnote{Originally we trained the network to also sample $\varphi_{M_3}$.} and we transform it into the distribution formed by the allowed parameter space of the various phases for a given pMSSM seed. We have opted to use Rational Quadratic Splines (RQS)~\cite{Rational_quadratic_spline:NEURIPS2019_7ac71d43} as our bijectors, having 7 of them in total. For each bijector we provide the parameters for the RQS by usage of a Multi-Layer-Perceptron~\cite{MLP:Cybenko1989} containing 2 layers and 512 intermediate features.\footnote{While the inclusion of Multi-Layer-Perceptrons in a RQS is typically understood as convention, we explicitly mention its inclusion here for clarity.} To condition the flow network on specific pMSSM model points, we provide the input parameters of the pMSSM seed as input to each of the Multi-Layer-Perceptrons, by appending it to the original input of the flow network.\\
For the learning process we use an ADAM optimizer~\cite{ADAM} with an initial learning rate of $5\cdot 10^{-4}$. We check the loss, for which we use the log likelihood, every 15 iterations during the learning process. If the loss has not improved by more than $0.001$ we decrease the learning rate by a factor of $0.1$. We use a total of 500 iterations.\\
\begin{figure}[t]
    \centering
    \includegraphics[width = .8\textwidth]{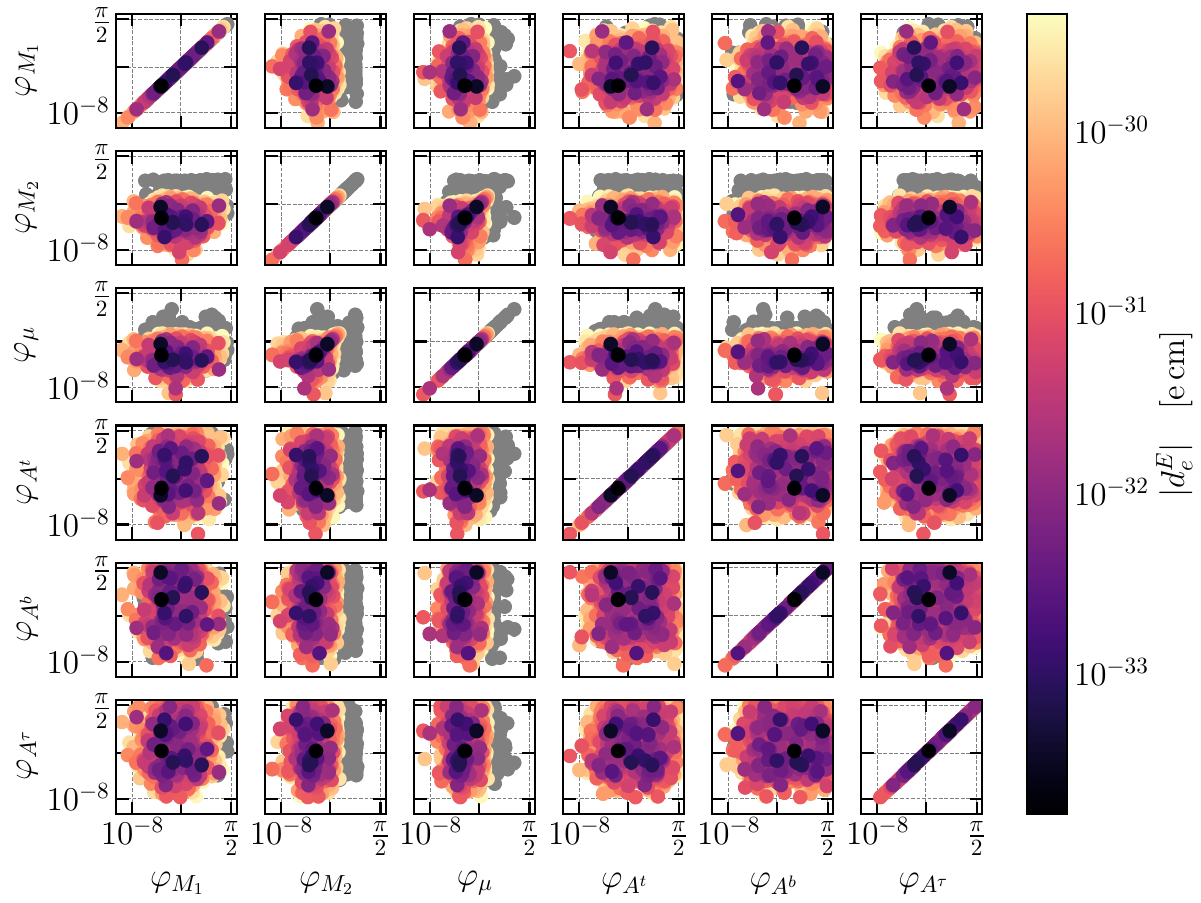}
    \caption{The sampled phases as provided by the trained flow network for a single pMSSM seed with the phases plotted against each other, with $\varphi_{M_1}$, $\varphi_{M_2}$, $\varphi_{\mu}$, $\varphi_{A^t}$, $\varphi_{A^b}$, $\varphi_{A^\tau}$ from top to bottom and left to right. The electron EDM is shown in colour, while the gray points are excluded by the limits specified in Section~\ref{subsec:Limits}. The model points have been sorted such that lower electron EDM values lie on top.}
    \label{fig:phase distribution}
\end{figure}
\begin{figure}[t]
    \centering
    \includegraphics[width=.8\textwidth]{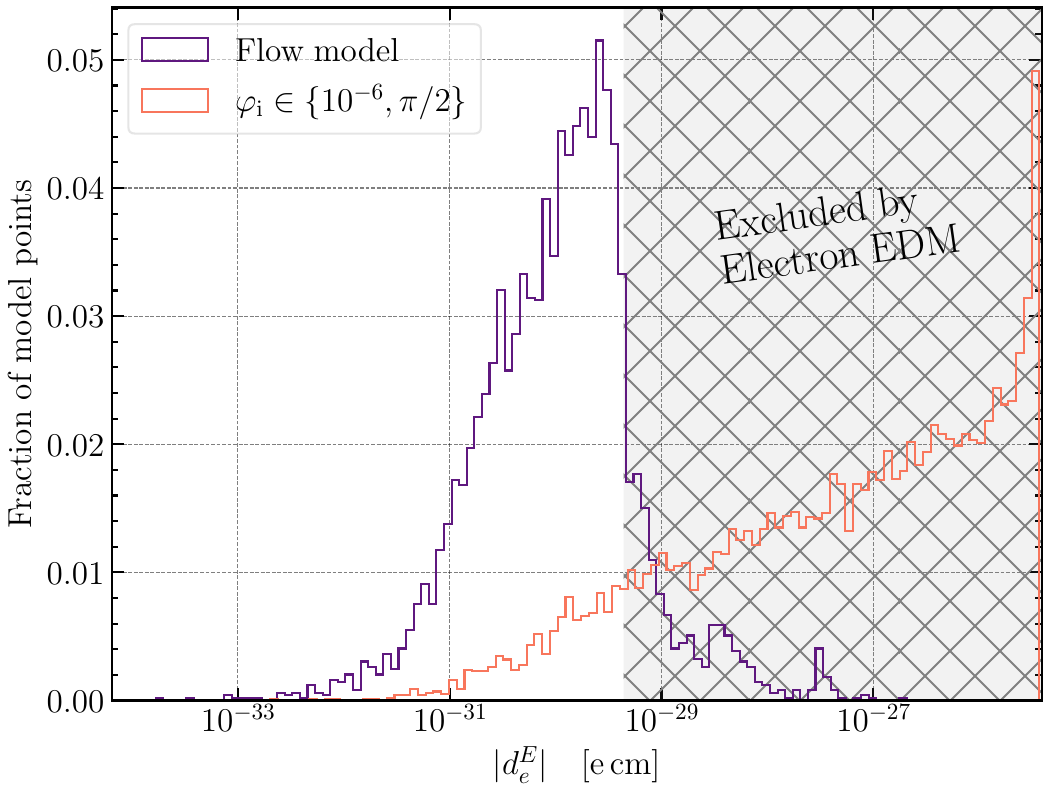}
    \caption{The distribution of electron EDM values for the samples provided by the trained normalizing flow model (purple) and samples obtained by independently and logarithmically sampling the six phases (red). All samples have been made with the same example pMSSM seed. The model points taken into consideration for the purple plot are the same as in Figure~\ref{fig:phase distribution}.}
    \label{fig:phase density}
\end{figure}
For our training data we use $5\cdot10^4$ different pMSSM model points, chosen to be representative of the original dataset, for which we have sampled $10^8$ sets of phases in total. For the training set, all phases are sampled independently on a logarithmically-uniform scale from $10^{-6}$ to $\pi/2$ and a minus sign is added to phases with a probability of 0.5 such that we effectively sample between $-\pi/2$ and $\pi/2$. We have sampled all phases independently to avoid introducing a bias into our training set.\footnote{The training set was sampled logarithmically, but we find that correcting for this bias provides no improvement on the accuracy of the model.} Notably, while we have sampled $10^8$ points, $1.8\cdot10^6$ remain after we cut for the electron EDM. By applying the transformation $\varphi \to -\varphi$ for all phases we enlarged our training data to $3.6\cdot10^6$ model points. This transformation leaves the electron EDM invariant. We find that our training set with $3.6\cdot10^6$ model points contains sufficient information to train the flow network to a satisfying degree; after training the flow network, $50\%-70\%$ of the samples it provides are non-excluded, depending on the input pMSSM model point. A more complete description of the model, training, and validation can be found in Ref.~\cite{Thesis_Dirren}.\\
We show the distribution of sampled phases made by the normalizing flow network for a typical pMSSM seed in Figure~\ref{fig:phase distribution}. Note that while we only show the samples for a single pMSSM seed, we observe little difference for the allowed ranges of the phases for other pMSSM seeds. The distribution of the electron EDM value for the same model points are shown in Figure~\ref{fig:phase density}, in addition to samples obtained by independently sampling the phases between $10^{-6}$ and $\pi/2$ for the same pMSSM seed. These figures only show the sampling for a single pMSSM point, the distributions can vary for different pMSSM seeds, however the images given are representative of the general trends. Most importantly, correlations between phases are not taken into account when independently sampling phases. The phases $\varphi_{M_2}$ and $\varphi_{\mu}$ are most sensitive to such correlations, as these phases are typically not allowed to simultaneously become sizable. The strong bias for high electron EDM values seen for independently sampled phases can be explained by the lack of tacking correlations into account, as multiple phases are typically large in such a scenario. Conversely, the trained flow network provides most of its samples on the exclusion boundary of the electron EDM, which is phenomenologically most interesting. 
\section{Results}
\label{sec:Results}
\subsection{The electron EDM limit and implications for the phases}
\label{subsec:The electron EDM limit and implications for the phases}
\begin{figure}[t]
    \centering
    \includegraphics[width = .8\linewidth]{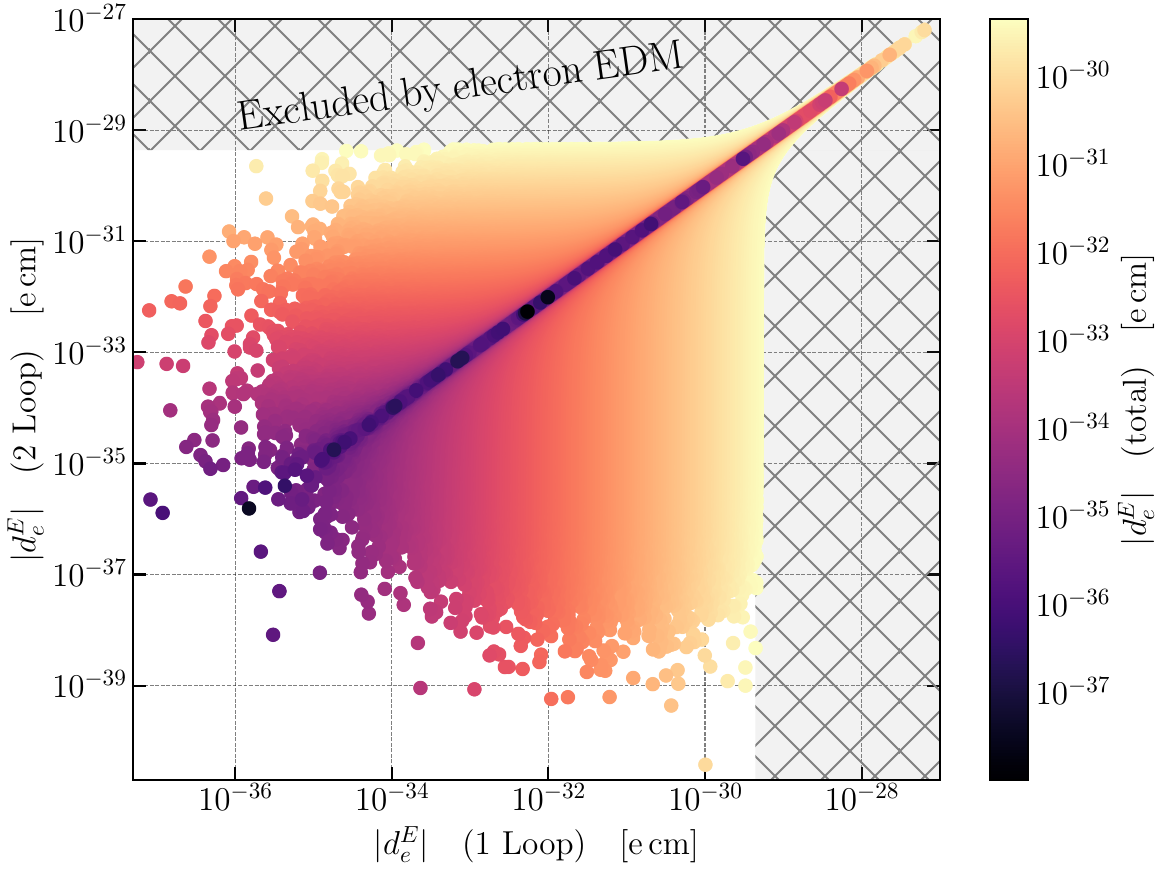}
    \caption{The absolute value of the electron electric dipole moment $|d^E_e|$ in e cm of the one-loop contribution against the two-loop contribution, with the total electron EDM in colour coding. The data has been sorted such that points with a lower electron EDM lie on top. All shown points obey the limits specified in Section~\ref{sec:Methodology}.}
    \label{fig:eEDM loop cancellations}
\end{figure}
It is well known that in certain scenarios the one and two-loop contributions of the muon anomalous magnetic moment can become of approximately equal size~\cite{gm2_loop1:JHEP, gm2_loop1_erratum:Cherchiglia2021, gm2_loop:PhysRevD.98.035001}. We find that this is also true for the electron EDM, as can be seen in Figure~\ref{fig:eEDM loop cancellations}. This is unsurprising, seeing as the anomalous magnetic moment and the electric dipole moment arise from taking the real and imaginary contributions of identical diagrams. Notably, cancellations between the one and two-loop contributions can become quite significant, reaching up to multiple orders of magnitude. We can find no clear correlation between values of the cpMSSM parameters and the presence of a cancellation between the one and two loop diagrams. We therefore suspect that there is no deeper physics underlying these cancellations aside from sufficient sampling, especially since these cancellations occur over two different orders in perturbation theory. Of course, the one and two loop contributions can also provide contributions with the same sign, which then increases the value of the electron EDM. Such model points are invisible in Figure~\ref{fig:eEDM loop cancellations}, since model points with lower electron EDM values lie on top. We find that such model points can always be found with sufficient sampling. It should be noted that the specific parameter values for which these cancellations occur might shift slightly, since some sub-dominant two-loop diagrams, and all third-or-higher-order corrections have not been taken into account. However, we still expect cancellations to occur when higher-precision computations are performed, seeing as we find a plethora of model points in the vicinity of the one-loop--two-loop-cancellation region.\\
Generally, we find for our spectra that phases associated with the electroweakinos, $\varphi_{M_1}$, $\varphi_{M_2}$, and $\varphi_{\mu}$, are more heavily constrained than those associated with the sfermions, $\varphi_{A^t}$, $\varphi_{A^b}$, $\varphi_{A^\tau}$. More specifically, we find that $\varphi_{M_2}$ and $\varphi_{\mu}$ are constrained up to a maximum of roughly $10^{-3}-10^{-2}$, depending on the masses and composition of the spectrum. Similarly, $\varphi_{M_1}$ is constrained up to $10^{-1}$, and $\varphi_{A^t}$, $\varphi_{A^b}$ and $\varphi_{A^\tau}$ are mostly unconstrained, with the exception of $\varphi_{A^t}$ and $\varphi_{A^\tau}$ for low masses. We find no appreciable difference for the ranges of the phases whether or not we include the model points in which a cancellation between the one- and two-loop contributions occurs. 
\begin{figure}[t]
    \centering
    \includegraphics[width = .8\textwidth]{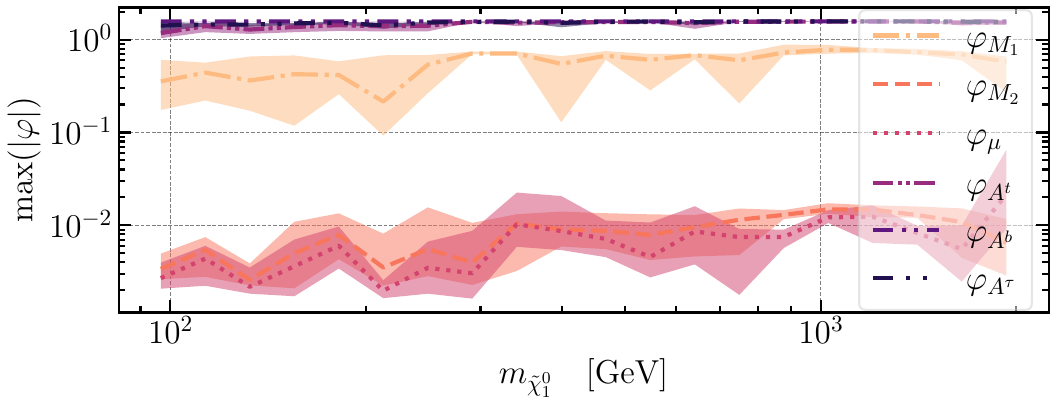}
    \caption{The maximum allowed values of the absolute values for $\varphi_{M_1}$, $\varphi_{M_2}$, $\varphi_{\mu}$, $\varphi_{A^t}$, $\varphi_{A^b}$, and $\varphi_{A^\tau}$. We have divided our complete dataset into five random batches, from which we compute the average maximum phases present, indicated with a dashed line, and the $1\sigma$ deviation, indicated by the shaded area.}
    \label{fig:max phases}
\end{figure}

\begin{figure}[t]
    \centering
    \includegraphics[width = .8\textwidth]{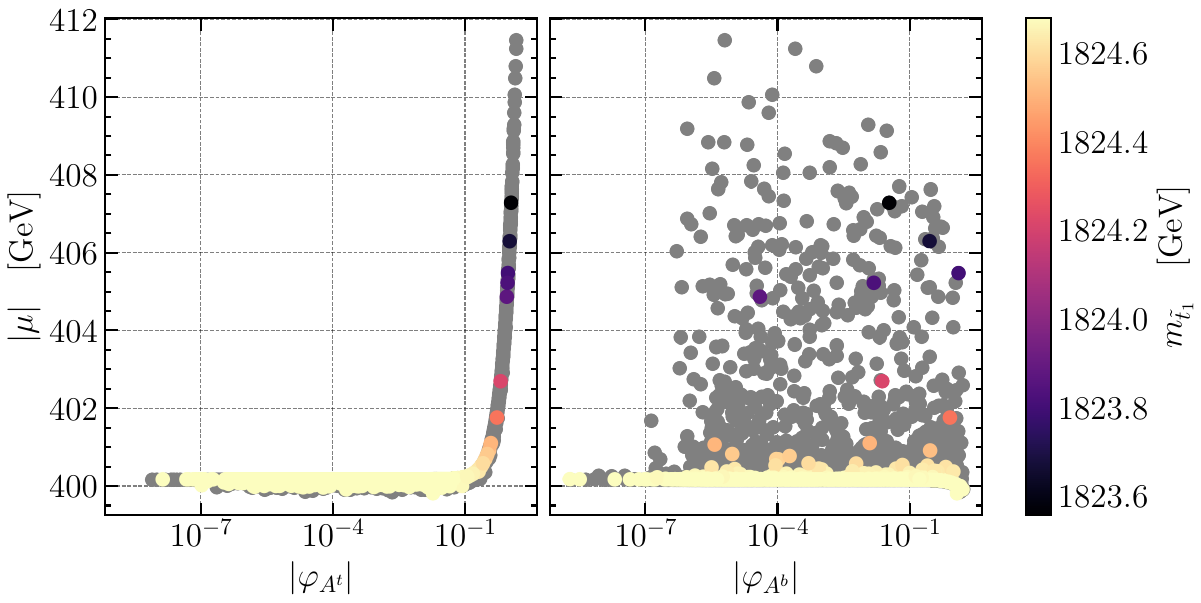}
    \caption{The absolute value of $\mu$ plotted against the absolute value of $\varphi_{A^t}$ (left) and $\varphi_{A^b}$ (right) with the mass of the lightest stop, $m_{\tilde{t}_1}$, as colour coding. All coloured points obey the limits specified in Section~\ref{sec:Methodology}, while the grey points are excluded. All points are generated from the same pMSSM seed.}
    \label{fig:phi_ATB against abs mu}
\end{figure}

\noindent Figure~\ref{fig:max phases} shows the maximum allowed absolute values of the six independent phases for a range of $\tilde{\chi}^0_1$ masses. For this figure we have divided our dataset into five random batches~\cite{Batching:10.1214/aos/1176344552}, from which we compute the average maximum value of each phase for each batch by taking the maximum value of the phase for each pMSSM seed point. The average of these maximum values is indicated by the solid line, and the $1\sigma$ deviation is shown by the shaded region.\\
We can see a slight increase for the various maximum values of $\varphi_{\mu}$ and $\varphi_{M_2}$ for increasing values of $m_{\tilde{\chi}^0_1}$. This is fairly unsurprising, seeing as the loop diagrams contributing to the electron EDM scale inversely with the masses of the relevant sparticles. Moreover, typically the wino and higgsino components couple more strongly to other sparticles than the bino component, which is reflected by the maximum allowed values of their phases, with $\varphi_{M_1}$ having a higher maximum by more than an order of magnitude than both $\varphi_{M_2}$ and $\varphi_{\mu}$ over the entire range of $m_{\tilde{\chi}^0_1}$. We emphasize that the irregular shape of both the average and shaded $1\sigma$ region are artefacts resulting from choosing the entries of the five batches at random, and are not the consequence of underlying physics.\\
Additionally, the electroweakinos appear at the one-loop level, thereby contributing at the lowest non-zero order to the electron EDM. Conversely, stops, sbottoms, and staus first appear at the two-loop level, as seen in Eq.~\eqref{eq:2_loop_eEDM_diagrams}. A priori we therefore expect these phases to have a smaller impact on the electron EDM, which is indeed what is seen when inspecting Figure~\ref{fig:max phases}. The phases $\varphi_{A^t}$, $\varphi_{A^b}$, and $\varphi_{A^\tau}$ are all allowed up to $\pi/2$, the upper limit of our sampling, across most values of $m_{\tilde{\chi}^0_1}$.\\
For $\varphi_{A^t}$ the situation is appreciably different. Since the value of $|\mu|$ is computed via the tadpole equations, as opposed to being fixed as an input parameter, its value is indirectly affected by phases. We find that $\varphi_{A^t}$ has the largest impact on $|\mu|$ of all the phases, followed by $\varphi_{A^b}$. In Figure~\ref{fig:phi_ATB against abs mu} we show an example model point to highlight the dependencies. In this figure, all data points are generated from the same pMSSM seed.

\noindent The change to the value of $|\mu|$ as a function of the phases arises from loop corrections to the Higgs potential, corresponding to the $V^{\text{corr}}$ terms in Eq.~\eqref{eq:tadpole eqations}, and more specifically the diagrams of Figures~\eqref{eq:one loop tadpole correction diagrams}~and~\eqref{eq: two loop tadpole corrections diagrams}. From these diagrams a dependence on the trilinear coupling can clearly be seen. However, these higher-order corrections additionally depend, in part, on the stop and sbottom masses, which are in turn affected by the phases of their trilinear couplings. Thus $\varphi_{A^t}$ and $\varphi_{A^b}$ influence $|\mu|$ in two ways: the first being the direct couplings in the loop-diagrams, and the second via the masses of the stops and sbottoms. Notably, the largest phase-induced changes to the stop and sbottom masses are given by loop corrections. At tree level $m^2_{\tilde{t}_{1,2}}$ and $m^2_{\tilde{b}_{1,2}}$ depend on the combinations $|A^t - \mu^*\cot(\beta)|$ and $|A^b - \mu^*\tan(\beta)|$ respectively, as can be inferred from Eq.~\eqref{eq:sfermion mass matrix}. However, the effect of $\varphi_\mu$ on the computed value of $|\mu|$ is significantly smaller than that of $\varphi_{A^t}$ and $\varphi_{A^b}$. This is because $\varphi_{\mu}$ only enters at tree-level in the mass matrices of the stops and sbottoms, while $\varphi_{A^t}$ and $\varphi^{A^b}$ have strong influences via loop corrections on the effective potential $V^{\text{corr}}$.\\
The main consequences of $|\mu|$ changing as a result of (mostly) $\varphi_{A^t}$, is that both the mass and composition of neutralinos and charginos can have a $\varphi_{A^t}$ dependence. Such effects are most pronounced for comparatively low values of $|\mu|$ and $m_{\tilde{t}_1}$, as the loop corrections of Eq.~\eqref{eq: two loop tadpole corrections diagrams} are then relatively large. Furthermore, a large $\varphi_{A^t}$ dependence occurs in various observables when the compositions of the neutralino or chargino in question are sensitive to small changes in $|\mu|$, e.g.~coannihilation regions for the DM relic density. As such $\varphi_{A^t}$ is constrained indirectly through its effect on the neutralino and chargino composition, as opposed to a more direct effect, as for example seen for $\varphi_{A^\tau}$.
\subsection{Direct detection}
\label{subsec:direct detection}
\begin{figure}[t]
    \centering
    \includegraphics[width = .8\textwidth]{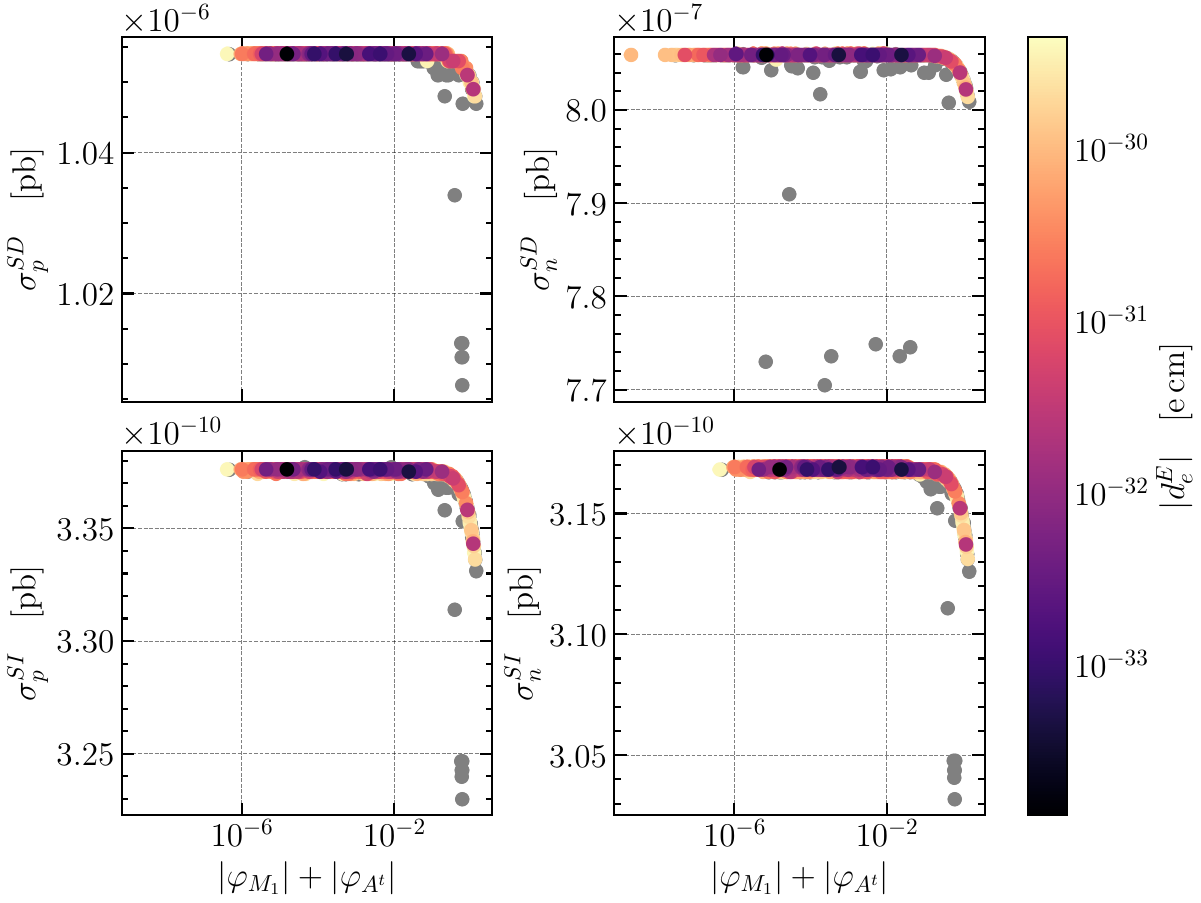}
    \caption{The spin-dependent (upper) and spin-independent (lower) cross sections of the DM candidate for the proton (left) and neutron (right) plotted against the sum of the absolute value of $\varphi_{M_1}$ and $\varphi_{A^t}$. We take the sum since these are the only two phases that we observe to have an effect on the direct detection cross sections, as such the sum captures the effect of both phases. The value of the electron EDM is shown in colour coding. All coloured points satisfy the limits given in Section~\ref{sec:Methodology}, while the grey points are excluded. The dark matter mass is $m_{\tilde{\chi}^0_1} = 1032$ GeV. All points have been created from the same pMSSM seed.}
    \label{fig:Direct detection}
\end{figure}
Seeing as the composition of the lightest neutralino is highly relevant for a number of DM observables, the addition of phases to a pMSSM model point can alter the associated value of the DM observable. The exact effect of a specific phase is in turn dependent on both the spectrum of the pMSSM seed and the DM observable. \\
The DM direct detection limits are one such observable. We can differentiate between the spin-dependent and spin-independent cross section; we find no appreciable difference between couplings to either a proton or neutron when inspecting the changes induced by non-zero phases. In Figure~\ref{fig:Direct detection} we show the four DM direct detection cross sections (spin-dependent proton and neutron cross section, and the spin-independent proton and neutron cross section) against the sum of the absolute values of $\varphi_{M_1}$ and $\varphi_{A^t}$. We take the sum since these are the only two phases we observe to provide an influence, not due to any correlation between $\varphi_{M_1}$ and $\varphi_{A^t}$ in DM direct detection cross sections. These two phases provide the largest influence, as we will specify later.\\
In general we find that the spin-dependent cross section is less affected by the introduction of phases than the spin-independent cross section. We can see why when inspecting their relevant couplings. The spin-dependent cross section is mostly mediated by the $Z$ boson, while the spin-independent cross section is mostly mediated by the Higgs bosons, with the lightest Higgs boson providing the largest contribution. The coupling of the lightest neutralino to the $Z$ boson is given by~\cite{Sparticles:drees2004theory}:
\begin{align}
    g_{\tilde{\chi}^0_1\tilde{\chi}^0_1Z} = 
    \adjustbox{valign = c}{
        \includegraphics{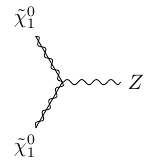}
    } = \frac{ig_2}{2c_W}\gamma^\mu \gamma^5 (|N_{13}|^2-|N_{14}|^2)\,, \label{eq: XXZ coupling}
\end{align}
where $c_W$ is the cosine of the Weinberg angle. From this we can see that only the absolute value of the neutralino mixing matrix $N$ appears. Thus, for the spin-dependent cross section only the absolute value of the composition is relevant; any real or imaginary components in the mixing matrix that are induced by the phases vanish from the total expression.\\
This is not the case for the coupling of the lightest neutralino to the Higgs bosons. We emphasize that the Higgs bosons in the cpMSSM are a mixture of the three mass eigenstates found in the pMSSM, as shown in Eq.~\eqref{eq: Higgs mixing}. The expression of the relevant coupling is~\cite{Sparticles:drees2004theory}:
\begin{align}
    &g_{\tilde{\chi}^0_1\tilde{\chi}^0_1h_j} = \adjustbox{valign = c}{
       \includegraphics{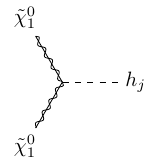}
    } = \sum_{\phi = \{h^0,H^0,A^0\}} ig_2R_{j\phi}\left( \mathfrak{Re}(Z_\phi) + i\gamma^5\mathfrak{Im}(Z_\phi) \right) \label{eq: XXh coupling} \\
    &Z_{h^0} = (N_{13}s_\alpha+N_{14}c_\alpha)(N_{12}-t_WN_{11})  \quad \quad \quad \quad Z_{H^0} = (-N_{13}c_\alpha+N_{14}s_\alpha)(N_{12}-t_WN_{11}) \nonumber \\
    &Z_{A^0} = (-iN_{13}s_\beta+iN_{14}c_\beta)(N_{12}-t_WN_{11}) \nonumber
\end{align}
Here $t_W$ is the tangent of the Weinberg angle and $s_\alpha$ $(c_\alpha)$ is the sine (cosine) of the CP-even Higgs mixing angle. From the expressions of Eq.\eqref{eq: XXh coupling} we can clearly see that both the real and imaginary part of the neutralino mixing matrix are relevant, as opposed to only the absolute contributions of the various components. Therefore the spin-independent cross section is, in general, more sensitive to the presence of phases than the spin-dependent cross section. \\
Furthermore, we find that $\varphi_{M_1}$ and $\varphi_{A^t}$ have the largest impact on $\sigma_{p/n}^{SD/SI}$. The effect of $\varphi_{A^t}$ is explained by its effect on $|\mu|$, while $\varphi_{M_1}$ influences the neutralino mixing matrix directly, by both being able to change the total size of $N_{i1}$, as well as the relative real and imaginary parts of $N_{i1}$. All other phases provide a negligible effect. In principle we expect a similar, if not larger, effect of $\varphi_{M_2}$ and $\varphi_{\mu}$ compared to $\varphi_{M_1}$. However, since these phases are constrained up to $\sim 10^{-2}$ by the limit of the electron EDM, their phases do not become large enough to significantly impact $\sigma_{p/n}^{SD/SI}$. Furthermore, current limits from DM direct detection experiments significantly constrain the amount of wino and higgsino component that are allowed in the lightest neutralino. Thus, since the lightest neutralino can have less wino and higgsino component due to these constraints, the values of $M_2$ and $\mu$ are constrained in the influence they have on the mixing of the lightest neutralino. This in turn limits the effect $\varphi_{M_2}$ and $\varphi_{\mu}$ on the direct detection cross sections. Thus current experiments mainly limit the effect of $\varphi_{M_2}$ and $\varphi_{\mu}$ by directly constraining their sizes through electron EDM measurements. Additionally DM direct detection and LHC searches limit the allowed amount of wino and higgsino component in the lightest neutralino, thereby indirectly suppressing the influence of $\varphi_{M_2}$ and $\varphi_{\mu}$. Notably the indirect effect is substantially weaker then direct electron EDM measurements. \\
\subsection{Indirect detection and Dark Matter relic density}
\label{subsec: sigmav and omegah}
We can further examine the velocity-weighted cross section $\langle \sigma v \rangle$ and the DM relic density $\Omega h^2$. We classify a model point according to its mediating particle featuring in the velocity-weighted cross section. Using this classification we differentiate between three different scenarios: those mediated by an $s$-channel SM boson, those mediated by a $t$-channel electroweakino, or those mediated by a $t$-channel sfermion. We can perform a similar categorization for the dominant contribution to the DM relic density, namely: a funnel scenario, electroweakino co-annihilation, and sfermion co-annihilation. We mention here explicitly that our dataset of pMSSM seeds contains no model points that have a $Z$ or $h_1$ funnel. This is due to the fact that the recent LZ limits have largely excluded those regions in the parameter space~\cite{No_funnel:Barman:2022jdg}. Unsurprisingly, the dependencies that we find for the velocity-weighted cross section categories are also inversely seen in relic density categories, while the diagrams providing these processes are not necessarily the same, such as for co-annihilation model points. The categories for the velocity-weighted cross section correspond to those of the DM relic density as follows: $s$-channel SM bosons corresponds to funnel regions, $t$-channel electroweakinos to electroweakino co-annihilation, and $t$-channel sfermions corresponds to sfermion co-annihilation.\\
\subsubsection{Electroweakinos}
\begin{figure}
    \centering
    \includegraphics[width = .8\textwidth]{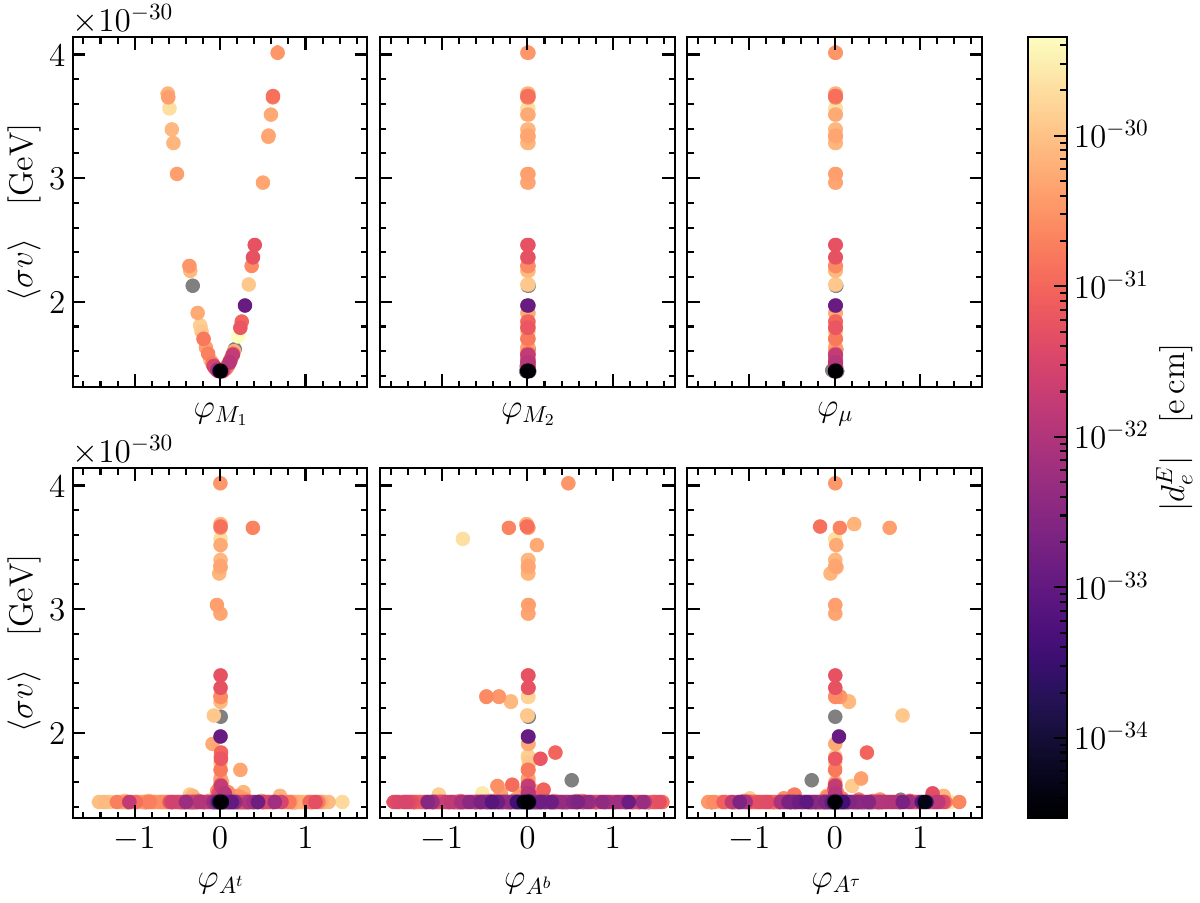}
    \includegraphics[width = .8\textwidth]{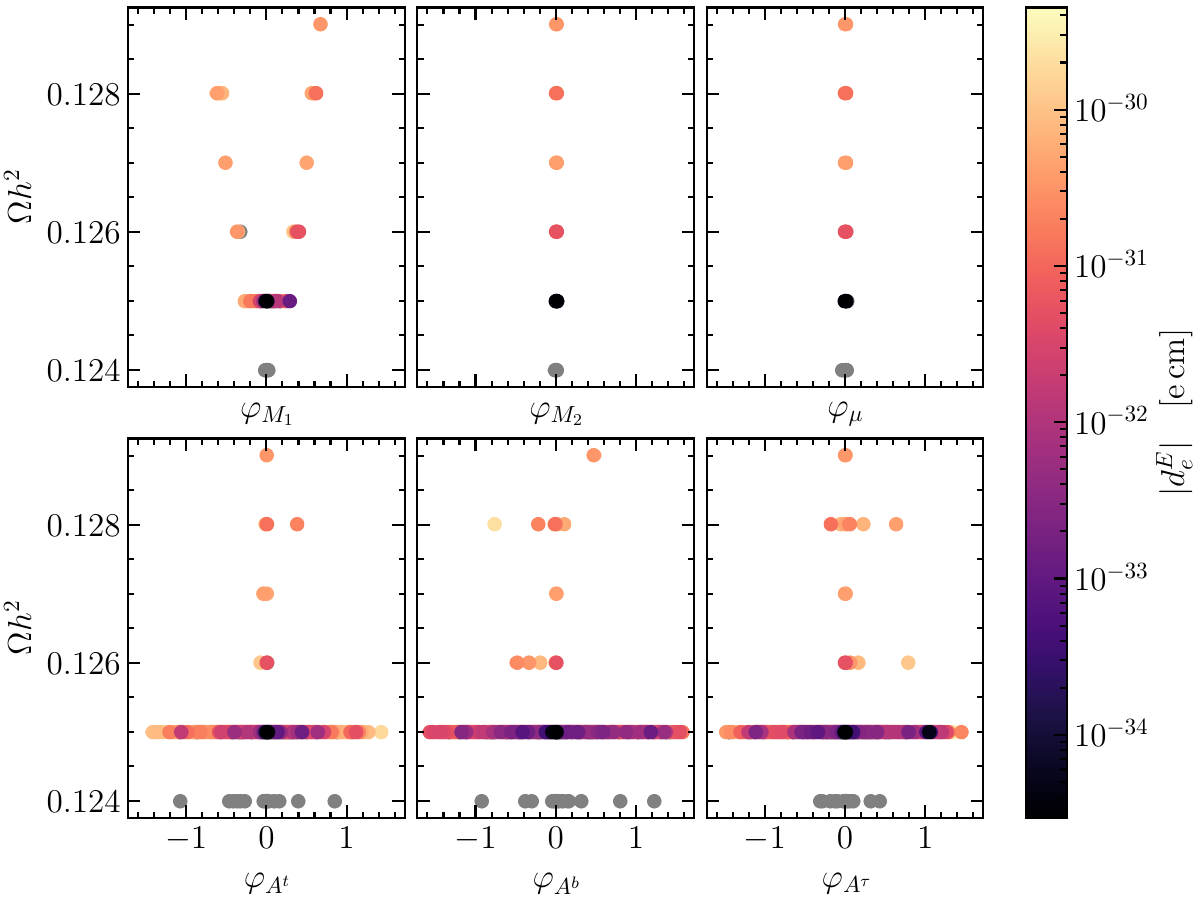}
    \caption{The velocity-weighted cross section (DM relic density) on top (bottom) against $\varphi_{M_1}$, $\varphi_{M_2}$, $\varphi_{\mu}$, $\varphi_{A^t}$, $\varphi_{A^b}$ and $\varphi_{A^\tau}$ from upper left to lower right respectively. All coloured points obey the limits of Section~\ref{sec:Methodology}, while the grey points are excluded. All six plots show the same points. Furthermore, the absolute value of the electron EDM is colour coded. All points are generated with the same pMSSM seed. The velocity-weighted cross section for all model points is dominated by a $t$-channel electroweakino, while the relic density is dominated by the electroweakino co-annihilation contribution. The DM mass is $m_{\tilde{\chi}^0_1} = 573~{\rm GeV}$.}
    \label{fig:sigmav omegah electroweakino}
\end{figure}
\begin{figure}
    \centering
    \includegraphics[width = .8\textwidth]{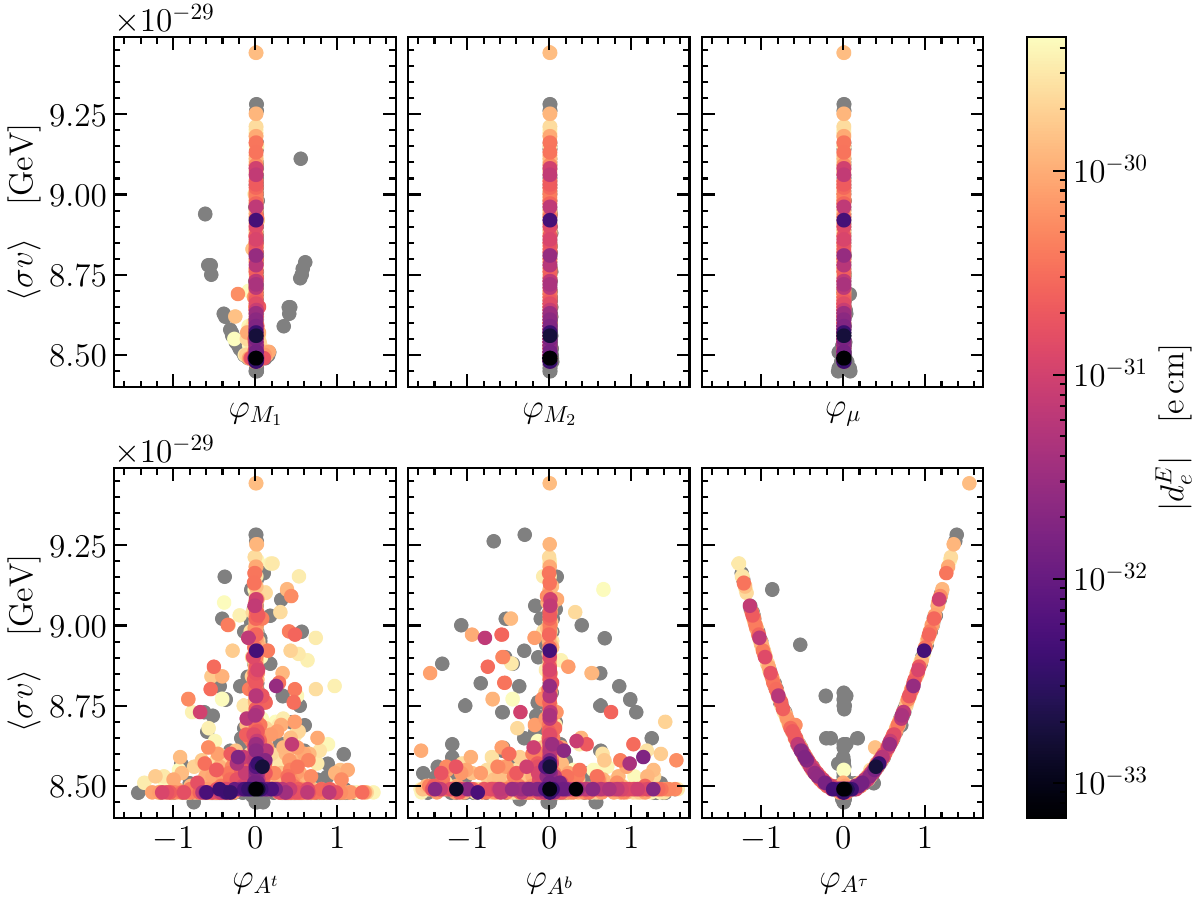}
    \includegraphics[width = .8\textwidth]{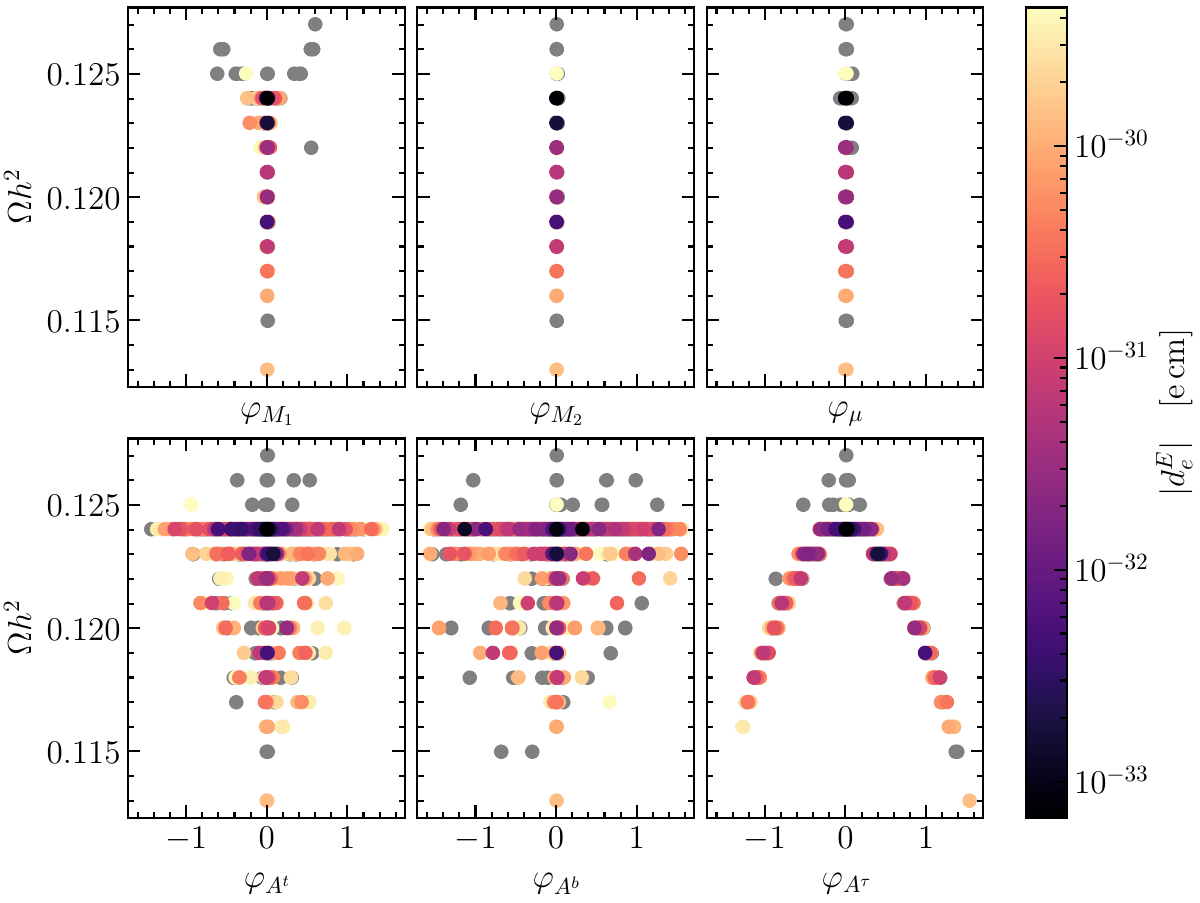}
    \caption{The same plots as in Figure~\ref{fig:sigmav omegah electroweakino}, except for a different pMSSM seed, such that the velocity-weighted cross section for all model points is dominantly provided by a $t$-channel stau, while the DM relic density is provided by stau co-annihilation.The DM mass is $m_{\tilde{\chi}^0_1} = 338~{\rm GeV}$.}
    \label{fig:sigmav omegah stau}
\end{figure}

In Figure~\ref{fig:sigmav omegah electroweakino} we show the effect of the various phases on $\langle \sigma v \rangle$ and $\Omega h^2$ respectively for when their sizes are dominated by electroweakino-driven processes. We also note that the fluctuations seen for $\varphi_{A^t}$, $\varphi_{A^b}$ and $\varphi_{A^\tau}$ arise when $\varphi_{M_1}$ also happens to be sizable. In general, when we notice fluctuations for a single phase for which a clear trend is not visible, then another phase that happens to provide the dominant contribution is sizable too. This can for example clearly be seen in Figure~\ref{fig:sigmav omegah stau} for $\varphi_{A^t}$ and $\varphi_{A^b}$. We additionally note that the gaps seen in the $\Omega h^2$ plots result from the finite resolution in {\tt MicroMegas}, these are also seen in later plots. We are currently unaware of any underlying reason why these gaps are of different sizes for the various plots.\\
In the electroweakino-driven category we find only a $\varphi_{M_1}$ dependence, as opposed to an additional $\varphi_{A^t}$ dependence, as would a priori be expected from its influence on $|\mu|$. We surmise that this is due to the fact that $\mu$ both appears in the neutralino and chargino mass matrix. Hence, a change in the value of $|\mu|$ will affect the composition of the various neutralinos and charginos approximately equally, because the masses of the two different electroweakinos involved in such a process must lie close together. Thus a change in $|\mu|$ due to $\varphi_{A^t}$ will not cause a significant change in the strength of the coupling involving two different electroweakinos.\\
The effect of $\varphi_{M_1}$ is slightly different. On the one hand, $\varphi_{A^t}$ influences the value of $|\mu|$, and consequently the total composition of the electroweakinos, but not necessarily the real and imaginary parts of the mixing matrices. On the contrary, $\varphi_{M_1}$ appears directly in the mass matrix of the neutralino, thereby also being able to change the relative sizes of the real and imaginary parts of the neutralino mixing matrix. As we have shown in Section~\ref{subsec:The electron EDM limit and implications for the phases}, the coupling of the neutralinos to the Higgs bosons is sensitive to such changes. In addition, since co-annihilation involves two different types of electroweakinos, the coupling to the vector bosons is no longer given by the absolute value of the mixing matrices, and therefore the coupling of the neutralinos and charginos to the vector bosons become sensitive to the ratio of the real and imaginary parts of the mixing matrices $N$, $\mathcal{V}$, and $\mathcal{U}$. The coupling of two different neutralinos to a $Z$ boson is given by:
\begin{align}
    g_{\tilde{\chi}^0_l\tilde{\chi}^0_kZ} = 
    \adjustbox{valign = c}{
        \includegraphics{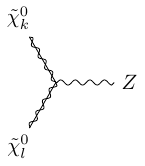}
    } = \frac{ig_2}{2c_W}\gamma^\mu \left[\gamma^5\mathfrak{Re}(N_{l3}N_{k3}^* - N_{l4}N_{k4}^*) -i \mathfrak{Im}(N_{l3}N_{k3}^* - N_{l4}N_{k4}^*)\right]\,, \label{eq: XlXkZ coupling}
\end{align}
while the coupling of two different charginos to a $Z$ boson is given by:
\begin{align}
    g_{\tilde{\chi}^\pm_l\tilde{\chi}^\pm_kZ} = 
    \adjustbox{valign = c}{
        \includegraphics{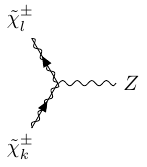}
    } = \begin{array}{l}\frac{ig_2}{2c_W}\gamma^\mu \big[-\mathcal{V}_{l1}\mathcal{V}^*_{k1} -\mathcal{U}_{l1}^*\mathcal{U}_{k1} - \frac{1}{2}(\mathcal{V}_{l2}\mathcal{V}^*_{k2} +\mathcal{U}_{l2}^*\mathcal{U}_{k2})  \\ 
    \phantom{\frac{ig_2}{2c_W}\gamma^\mu \big[}+ (\mathcal{V}_{l1}\mathcal{V}^*_{k1} -\mathcal{U}_{l1}^*\mathcal{U}_{k1} + \frac{1}{2}(\mathcal{V}_{l2}\mathcal{V}^*_{k2} -\mathcal{U}_{l2}^*\mathcal{U}_{k2}))\gamma^5 + 2\delta_{lk} s_W^2\big]\end{array}\,. \label{eq: ClCk coupling}
\end{align}
When $l=k$ we can see for both Eq.~\eqref{eq: XlXkZ coupling} and Eq.~\eqref{eq: ClCk coupling} that only the absolute values of the mixing matrices are relevant. The coupling of a neutralino-chargino pair to a $W^\pm$ boson never involves the absolute values of the mixing matrices.
\subsubsection{$h_2$ funnel}
\begin{figure}
    \centering
    \includegraphics[width = .8\textwidth]{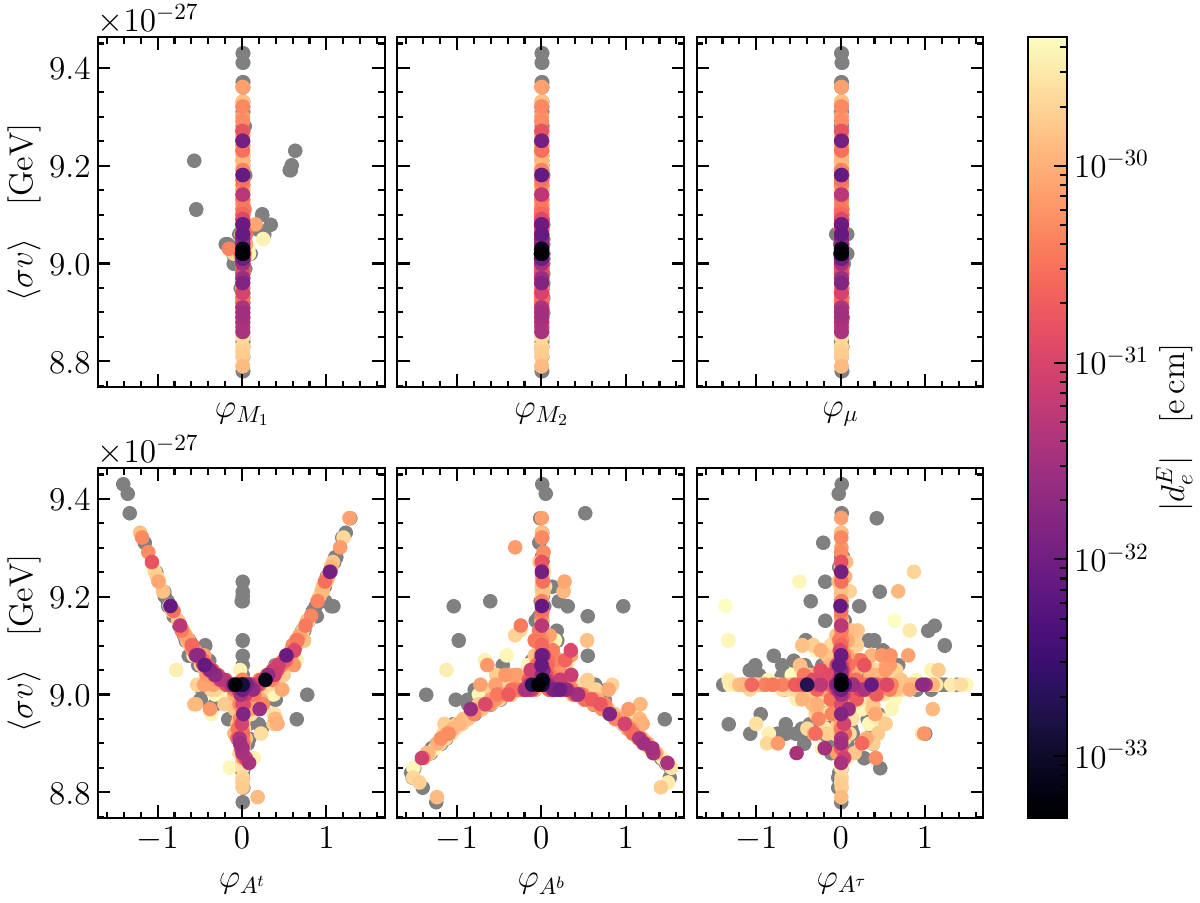}
    \includegraphics[width = .8\textwidth]{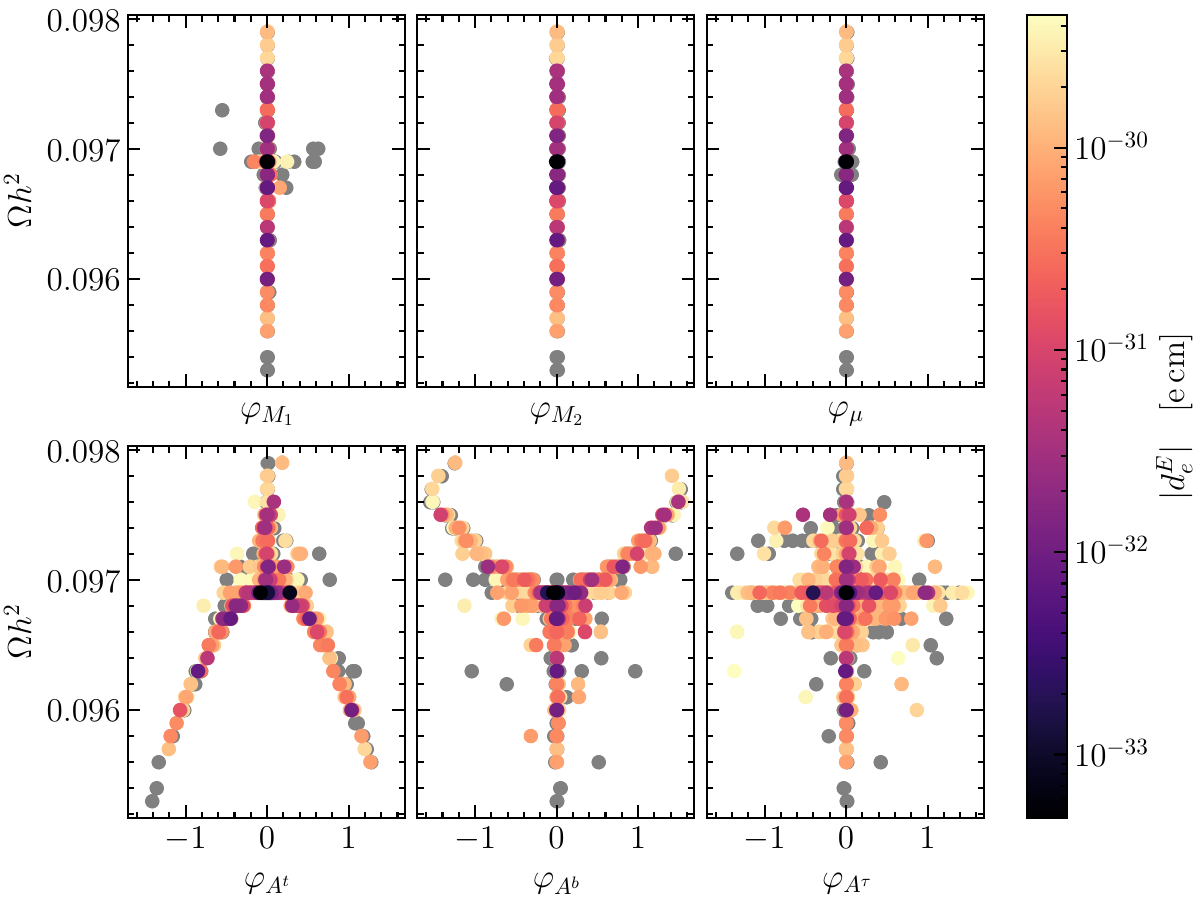}
    \caption{The same plots as in Figure~\ref{fig:sigmav omegah electroweakino}, except for a different pMSSM seed, such that the velocity-weighted cross section for all model points is dominantly provided by a mediating $h_2$ in an $s$-channel, while the DM relic density is provided by a $h_2$ funnel. The DM mass is $m_{\tilde{\chi}^0_1} = 880~{\rm GeV}$.}
    \label{fig:sigmav omegah boson}
\end{figure}
\begin{figure}[h]
    \centering
    \includegraphics[width = \textwidth]{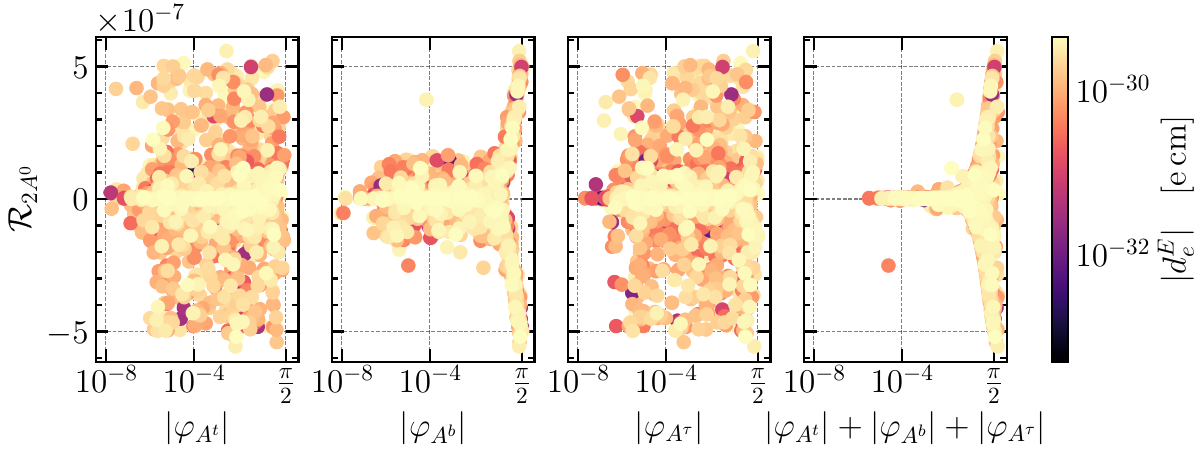}
    \caption{The CP-odd Higgs mixing parameter for $h_2$ against the absolute values of $\varphi_{A^t}$, $\varphi_{A^b}$, $\varphi_{A^\tau}$, and the sum of their absolute values. All points have the same pMSSM seed. The points shown are the same as in Figure~\ref{fig:sigmav omegah boson}.}
    \label{fig:Higgs mixing}
\end{figure}
In Figure~\ref{fig:sigmav omegah boson} we show both the effect of phases on $\langle \sigma v \rangle$ and $\Omega h^2$ for a $h_2$-mediated process. Here we see an influence of both $\varphi_{A^t}$ and $\varphi_{A^b}$. Since the coupling of the lightest neutralino to $h_2$ scales with the composition of $\tilde{\chi}^0_1$, the effect of $\varphi_{A^t}$ is readily explained. The effect of $\varphi_{A^b}$ arises due to the Higgs mixing matrix $\mathcal{R}$, as defined in Eq.~\eqref{eq: Higgs mixing}. Since $H^0$ and $A^0$ couple more strongly to down-type particles, by a factor of $\tan^2(\beta)$ as compared to up-type particles, the bottom squark has a relatively strong influence as compared to the top squark on the mixing of CP-odd components in $h_2$. The mixing of the CP-even and CP-odd mass eigenstates is then dominated by the trilinear couplings. In Figure~\ref{fig:Higgs mixing} we show the dependency of the CP-odd component of $h_2$ in terms of the absolute values of the phases of the trilinear couplings. In fact, for the sizes of $\mathcal{R}_{1A^0}$, $\mathcal{R}_{2A^0}$, $\mathcal{R}_{3h^0}$, and $\mathcal{R}_{3H^0}$ we observe similar dependencies for $h_1$, $h_2$ and $h_3$ on the phases. The points shown in Figure~\ref{fig:Higgs mixing} are the same as those in Figure~\ref{fig:sigmav omegah boson}. From this we can also infer that the trilinear couplings have the largest influence on the Higgs mixing matrix $\mathcal{R}$.\\
\subsubsection{Staus}
Lastly, for a DM relic density driven by sfermion co-annihilation and sfermion-driven $\langle \sigma v \rangle$, we find that the phase associated with said sfermion is most relevant. For example, in stau-coannihilation $\varphi_{A^\tau}$ has the dominant effect, as shown in Figure~\ref{fig:sigmav omegah stau}. Such an effect is of course expected, especially when inspecting the neutralino-tau-stau coupling\footnote{The diagram in which the $\tau$ is incoming and $\tilde{\tau}$ is outgoing only differs by a complex conjugate on all mixing matrices, and changing $P_R \leftrightarrow P_L$}~\cite{Sparticles:drees2004theory}:
\begin{flalign}
    g_{\tilde{\tau}_j \tau\tilde{\chi}^0_i} = \adjustbox{valign = c}{
        \includegraphics{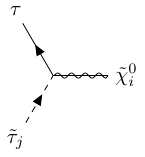}
    } \begin{array}{r} = \frac{ig_2}{\sqrt{2}}\big(((N_{i2}+t_WN_{i1})P_R-h^\tau N_{i3}^*P_L)X^{\tilde{\tau}*}_{jL}\phantom{\big)} \nonumber \\
    -(2t_W N_{i1}^*P_L+h^\tau N_{i3}P_R)X^{\tilde{\tau}*}_{jR}\big)\end{array}\,. \label{eq: tau stau neutralino1 1}
\end{flalign}
Here $h^\tau = 2m_\tau/(g_2 v_d)$ and $P_{L/R} = 1/2\,(1\,{\!-/+}\,\gamma^5)$. This coupling contains a dependency on not only all neutralino components, but also the stau mixing matrix $X^{\tilde{\tau}}$. The latter is clearly influenced by $\varphi_{A^\tau}$, as seen in Eq.~\eqref{eq:sfermion mass matrix}. A subdominant effect of $\varphi_{M_1}$ can also be present, provided that large values of $\varphi_{M_1}$ are not excluded by the electron EDM limit. The $\varphi_{M_1}$ dependence can again be explained by the dependency of both the composition of the neutralino and the relative real and imaginary components of $N_{1i}$ in the neutralino-tau-stau coupling. While not shown, model points in which stop or sbottom co-annihilation occurs show a similar dependence on $\varphi_{A^t}$ and $\varphi_{A^b}$ respectively.

\subsection{General trends}
We finish our discussion of the results by noting that for our model points in general we find that $\langle \sigma v \rangle$ typically increases as a result of adding phases, while $\Omega h^2$ typically decreases. Furthermore, we find that when adding phases $\sigma_{p/n}^{SD}$ typically changes less than $\sigma_{p/n}^{SI}$. Regarding the changes caused by adding phases, we notice little difference between couplings to the proton or neutron for both the spin-dependent and spin-independent case. This is due to the fact that the difference for these couplings is how the $Z$ and Higgs bosons couple differently to the up and down quarks, but the effects of non-zero phases are seen in the coupling of the $Z$ and Higgs bosons to the lightest neutralino. Notably, when considering the maximum change from the pMSSM seed, the DM relic density changes the least of all considered DM observables. Lastly, while we have shown multiple figures in which only a single pMSSM seed was used, the results that have been discussed are seen in general for our spectra.\footnote{The data can be found in \href{https://zenodo.org/doi/10.5281/zenodo.10606167}{this} Zenodo repository.}\\
We additionally find that collider and direct detection experiments provide an influence on the size of the allowed phases in the cpMSSM. Naturally, the increased mass limits from the various LHC experiments on the various sparticles push down the contributions to the electron EDM via mass suppression. Especially limits on the stop, sbottom and stau mass influence the size of the Barr-Zee diagrams. For the one loop diagrams we find that the composition of the neutralinos are most important. Here direct detection experiments provide strong bounds on the allowed wino and higgsino components in the lightest neutralino, which in turn affects the size of the one-loop diagrams for comparable phases.

\section{Conclusion}
\label{sec:Conclusion}
In this paper we have studied the impact of CP-violating phases on a selection of DM observables in the framework of the cpMSSM.
This supersymmetric model has the same particle content as the pMSSM, but 6 additional independent phases are added to the bino, wino and higgsino mass parameters $M_1$, $M_2$ and $\mu$, and the trilinear couplings of the third generation sparticles $A^t$, $A^{\tau}$ and $A^b$. We constrain the pMSSM parameter space by setting the values of the first/second generation squarks to $2$~TeV, and the gluino mass parameter to $2.5$~TeV. We then select pMSSM model points that are not excluded by LHC and DM measurements, and subsequently train a normalizing flow network to sample phases for these individual points. We find that employing such a scanning procedure allows us to efficiently sample cpMSSM points near the electron EDM exclusion boundary (compared to sampling the phases independently and uniformly or logarithmically). \\
Phases of the cpMSSM are generally assumed to be small due to the non-observation of the electron EDM, which places an upper bound of $4.1\cdot 10^{-30}\,\,{\rm e\, cm}$. We find that the absolute size of $\varphi_{M_2}$ and $\varphi_{\mu}$ are indeed constrained to an approximate maximum absolute value of $10^{-2}$, and $\varphi_{M_1}$ is limited to $10^{-1}$ by the electron EDM measurements (assuming electroweakino masses of maximally $\mathcal{O}(1)$~TeV). While we have used two different datasets, one optimised for explaining the $(g-2)_\mu$ discrepancy and one with a flat prior over the pMSSM parameter space, we find no appreciable difference between the allowed phases for these two datasets. We have not optimised our search to find model points with large phases, but rather focussed on the phenomenology for a variety of typical cpMSSM model points with sparticle masses between $\mathcal{O}(100)$~GeV and $\mathcal{O}(10)$~TeV. The fact that $\varphi_{M_2}$ and $\varphi_{\mu}$ are most constrained is not surprising, as it is a direct consequence of the most dominant diagrams that contribute to the electron EDM, which typically involve an exchange of an electroweakino. However, contrary to common belief, we find that the phases of the trilinear couplings $\varphi_{A^t}$, $\varphi_{A^b}$ and $\varphi_{A^\tau}$, are largely unconstrained by electron EDM measurements and even $\varphi_{M_1}$ is allowed to be sizable.\\
We studied the impact of these phases on the DM direct detection cross sections, indirect DM annihilation cross section, and the DM relic density. As $|\varphi_{M_2}|$ and $|\varphi_{\mu}|$ are already quite constrained, their value does not impact the above-mentioned observables. However, we find that $\varphi_{M_1}$ and $\varphi_{A^t}$ mostly impact the spin-independent direct detection cross section. The former impacts the neutralino mixing matrix directly, whereas $\varphi_{A^t}$ indirectly affects the value of $|\mu|$. The latter is a consequence from constraining $|\mu|$ via the tadpole equations, where $\varphi_{A^t}$ affects the size of the radiative corrections to the Higgs potential. Diagrams that contribute to $\sigma_{p,n}^{SI}$ always feature an incident DM particle that is scattered off a quark, thereby limiting the set of diagrams to those that involve an $s$- and $t$-channel squark or $t$-channel Higgs boson exchange. This is in contrast with $\langle \sigma v \rangle$ and $\Omega h^2$, where a larger variety of scatterings can occur. For the electroweakino-dominated channels we find that $\varphi_{M_1}$ affects the value of $\langle \sigma v \rangle$ and $\Omega h^2$. Unsurprisingly, we find that $\varphi_{A^\tau}$ provides a significant influence for channels involving staus. For channels that feature an $h_2$ funnel, we additionally see that $\varphi_{A^t}$ and $\varphi_{A^b}$ affect these DM observables. This is because the Higgs mixing matrix $\mathcal{R}$ depends on the phases of the trilinear couplings, with $\varphi_{A^b}$ greatly affecting the mixing of the CP-even and CP-odd components of the Higgs mass eigenstates. The typical values for $\langle \sigma v \rangle$ that feature for our cpMSSM model points are typically too small to be detected by DM indirect detection experiments. However,  $\Omega h^2$ is constrained to be within $0.09-0.15$ in our study, and for $\sigma^{\rm SI}$ we probe points that lie on the boundary of exclusion. For such points, it is important to take into account the presence of a possible cpMSSM phase, as such phases can affect the values of these observables to the extent that they are either pushed into the observable region, or inside the excluded region. \\
In this paper we have only studied the impact of phases in the cpMSSM on DM observables. It would be of interest to also study their collider phenomenology, especially in the context of searches for CP violation. However, adding such an analysis requires both properly integrating dedicated software into our computation chain and performing an entirely new analysis. Especially carefully implementing LHC limits, as opposed to cuts on sparticle masses as we have done here, requires a resampling of the cpMSSM parameter space near the exclusion boundary, since regions we consider to be excluded might in actuality not be so, most notably in the coloured sector. As such we leave performing such an analysis to future work. \\
Future particle-physics experiments will further probe the allowed parameter space. Both DM direct detection experiments and LHC searches will search for the presence of new (SUSY) particles, while electron EDM measurements are expected to provide information regarding the size of phases in the cpMSSM. A combined analysis of multiple experiments is key to fully probe the cpMSSM parameter space. We stress that complex phases can move pMSSM points that lie on the boundary of discovery to either the excluded or non-excluded region, meaning that these phases cannot be ignored.

\Acknowledgements 
MS would like to thank Rob Timmermans for his help and many useful discussions, and additionally the hospitality of the Oxford University.
\FloatBarrier
\bibliography{main}
\bibliographystyle{JHEP}
\end{document}